\documentclass[twocolumn,aps,prx,superscriptaddress,nofootinbib,floatfix]{revtex4-2}

\usepackage{graphicx}
\usepackage[dvipsnames,svgnames]{xcolor}
\usepackage{bm,bbm}
\usepackage{braket}
\usepackage{amsthm,amsmath,amssymb}
\usepackage{enumerate}
\usepackage{booktabs,multirow}
\usepackage{mathtools,bbding}
\usepackage[percent]{overpic}
\usepackage{microtype}
\usepackage{array,adjustbox,afterpage}
\usepackage[T1]{fontenc}
\usepackage[utf8]{inputenc}
\usepackage{tikz}
\usetikzlibrary{calc}
\usepackage[english]{babel}
\usepackage{newtxtext,newtxmath}
\usepackage{dsfont}

\usepackage[linesnumbered,ruled]{algorithm2e}

\makeatletter
\renewcommand{\fnum@figure}{FIG. \thefigure}
\makeatother
\makeatletter
\renewcommand{\fnum@table}{TABLE \thetable}
\makeatother

\usepackage[bookmarks=false, colorlinks=true, linkcolor=blue, citecolor=purple, urlcolor=purple]{hyperref}
\usepackage{cleveref}
\usepackage[normalem]{ulem}

\newif\ifshowedits
\showeditsfalse

\ifshowedits

  \newcommand{\LGdel}[1]{\textcolor{Blue}{\sout{#1}}}
  \newcommand{\BAdel}[1]{\textcolor{ForestGreen}{\sout{#1}}}
  \newcommand{\YGdel}[1]{\textcolor{OrangeRed}{\sout{#1}}}
  \newcommand{\OPdel}[1]{\textcolor{Purple}{\sout{#1}}}
  \newcommand{\IKdel}[1]{\textcolor{Magenta}{\sout{#1}}}
  \newcommand{\OMdel}[1]{\textcolor{Teal}{\sout{#1}}}
  \newcommand{\FPdel}[1]{\textcolor{Brown}{\sout{#1}}}
  
  \newcommand{\TCdel}[1]{\textcolor{Red}{\sout{#1}}}
\else

  \newcommand{\LGdel}[1]{}
  \newcommand{\BAdel}[1]{}
  \newcommand{\YGdel}[1]{}
  \newcommand{\OPdel}[1]{}
  \newcommand{\IKdel}[1]{}
  \newcommand{\OMdel}[1]{}
  \newcommand{\FPdel}[1]{}
  \newcommand{\TCdel}[1]{}
\fi

\Crefname{subfigures}{figure}{figures}
\Crefname{subfigures}{Figure}{Figures}

\newcommand{\mapcell}[2]{\parbox[t]{#1}{\raggedright #2}}

\hypersetup{
    bookmarksnumbered=false,
    breaklinks=true,
    unicode=false,
    pdfstartview={FitH},
    pdftitle={},
    pdfnewwindow=true,
}

\begin{document}

\title{Quantum in Biology, Quantum for Biology, and Biology for Quantum: Mapping the Evidence and the Road Ahead}

\author{Lea Gassab}
\affiliation{Departments of Biology, Chemistry, and Physics \& Astronomy, Waterloo Institute for Nanotechnology, University of Waterloo, Waterloo, ON, Canada}

\author{Betony Adams}
\affiliation{Department of Physics, Stellenbosch University, Stellenbosch 7600, South Africa}
\affiliation{School of Data Science and Computational Thinking, Stellenbosch University, Stellenbosch, South Africa}
\affiliation{National Institute for Theoretical and Computational Sciences, Stellenbosch, South Africa}
\affiliation{The Guy Foundation, Dorset, United Kingdom}

\author{Yashine H. Goolam Hossen}
\affiliation{Departments of Biology and Physics \& Astronomy, Waterloo Institute for Nanotechnology, University of Waterloo, Waterloo, ON, Canada}
\affiliation{Waterloo Institute for Complexity and Innovation, University of Waterloo, Waterloo, ON, Canada}

\author{Abbas Hassasfar}
\affiliation{Department of Physics, Stellenbosch University, Stellenbosch 7600, South Africa}
\affiliation{National Institute for Theoretical and Computational Sciences, Stellenbosch, South Africa}

\author{Onur Pusuluk}
\affiliation{Faculty of Engineering and Natural Sciences, Kadir Has University, Fatih 34083, Istanbul, T\"urkiye}

\author{Nirosha J. Murugan}
\affiliation{Department of Health Sciences, Wilfrid Laurier University, Waterloo, ON N2L 3C5, Canada}
\affiliation{Allen Discovery Center, Tufts University, Medford, MA 02155, USA}

\author{Iannis K. Kominis}
\affiliation{School of Science, Zhejiang University of Science and Technology, Hangzhou 310023, China}
\affiliation{Department of Physics and Institute of Theoretical and Computational Physics, University of Crete, Heraklion 70013, Greece}

\author{\"Ozg\"ur E. M\"ustecapl{\i}o\u{g}lu}
\affiliation{Department of Physics, Ko\c{c} University, 34450 Sar{\i}yer, Istanbul, T\"urkiye}
\affiliation{T\"UB\.{I}TAK Research Institute for Fundamental Sciences (TBAE), 41470 Gebze, T\"urkiye}

\author{Francesco Petruccione}
\affiliation{School of Data Science and Computational Thinking, Stellenbosch University, Stellenbosch, South Africa}
\affiliation{National Institute for Theoretical and Computational Sciences, Stellenbosch, South Africa}

\author{Travis J. A. Craddock}
\affiliation{Departments of Biology and Physics \& Astronomy, Waterloo Institute for Nanotechnology, University of Waterloo, Waterloo, ON, Canada}

\begin{abstract}
Quantum science and biology now intersect in three complementary directions: quantum in biology, quantum for biology, and biology for quantum. This review provides a structured narrative evidence map of that interface rather than an exhaustive catalogue or formal systematic review. For each topic, we ask what the mechanistic or technological claim is, which quantum resource is invoked, what the strongest experiments and models establish, which classical alternatives or engineering confounds remain competitive, and what decisive tests or benchmarks would most strongly change confidence. The most mature quantum-in-biology cases remain mechanistically constrained tunneling in some enzymatic hydrogen-transfer reactions and radical-pair spin chemistry as a viable framework for magnetoreception, whereas several higher-visibility topics remain suggestive but unresolved under physiological conditions. In quantum for biology, the central issue is whether quantum-enabled tools improve biological inference relative to strong classical baselines under realistic calibration, dose, throughput, and uncertainty constraints. In biology for quantum, the strongest claims arise when biomolecular structure or self-assembly measurably improves fabrication, integration, or robustness in quantum devices. Summary tables in the Appendix provide a compact cross-map view of the current evidence, major confounds, and the experiments or benchmarks most likely to discriminate between competing explanations.
\end{abstract}

\maketitle
\tableofcontents


\section{Introduction}

Quantum mechanics is indispensable for chemistry and therefore for biology at the molecular scale. The stricter question posed by quantum biology, however, is not whether biology depends on quantum mechanics in principle, but whether particular quantum resources, such as nuclear or electron tunneling, vibronic mixing, or spin-dependent radical-pair dynamics, are necessary to explain biological observables under realistic conditions, rather than merely providing a more microscopic description of processes that are already adequately captured by classical or semiclassical models \cite{Lambert2013,Marais2018,ball2011physics,ScholesFleming2026,Babcock2025PhysicalPrinciples}.

This question has a long intellectual history. Schr\"odinger helped frame heredity and biological order as problems for fundamental physics, and later work sought more specific mechanisms by which quantum effects might matter in living systems \cite{schrodinger1944life}. Several canonical examples illustrate both the promise and the difficulty of such claims. L\"owdin proposed that proton transfer in DNA hydrogen bonds could generate rare tautomers and thereby influence mutation \cite{loewdin1963proton}. Enzymatic hydrogen transfer has become a central case in which tunneling contributions can be supported by convergent evidence from reaction kinetics, including kinetic isotope effects and their temperature dependences \cite{klinman2013hydrogen}. Studies of photosynthetic complexes have revealed ultrafast oscillatory signals that can be described using coherent or vibronically mixed excitonic dynamics, although their functional significance under physiological conditions remains debated because vibrational and ensemble effects can mimic signatures otherwise attributed to electronic coherence \cite{Engel2007,Panitchayangkoon2010,Collini2010,runeson2022explaining}. Radical-pair spin chemistry likewise provides a concrete framework by which weak magnetic fields could influence reaction yields and potentially support magnetoreception, although important uncertainties still exist regarding the relevant radicals in vivo and the pathways by which spin-dependent chemistry is transduced into biological response \cite{Ritz2000,Cai2010,cai2013chemical}.

The relationship between quantum science and biology also extends beyond mechanistic questions within living systems. Quantum technologies are increasingly being developed as tools for biological measurement, sensing, imaging, molecular simulation, and data analysis, where the relevant question is whether they improve biological inference or experimental capability relative to strong classical baselines under realistic constraints \cite{Schirhagl2014,degen2017quantum,moreau2019imaging,Cao2020}. Conversely, biology may also inform quantum science through self-assembly, molecular organization, biomaterial platforms, and design principles that suggest new strategies for fabrication, integration, and quantum device design \cite{kuzyk2018dnaorigami,Creatore2013,feder2025fluorescentqubit}.


\section{How to read this review as a map}

\begin{figure*}[t]
    \centering
    \includegraphics[width=0.9\textwidth]{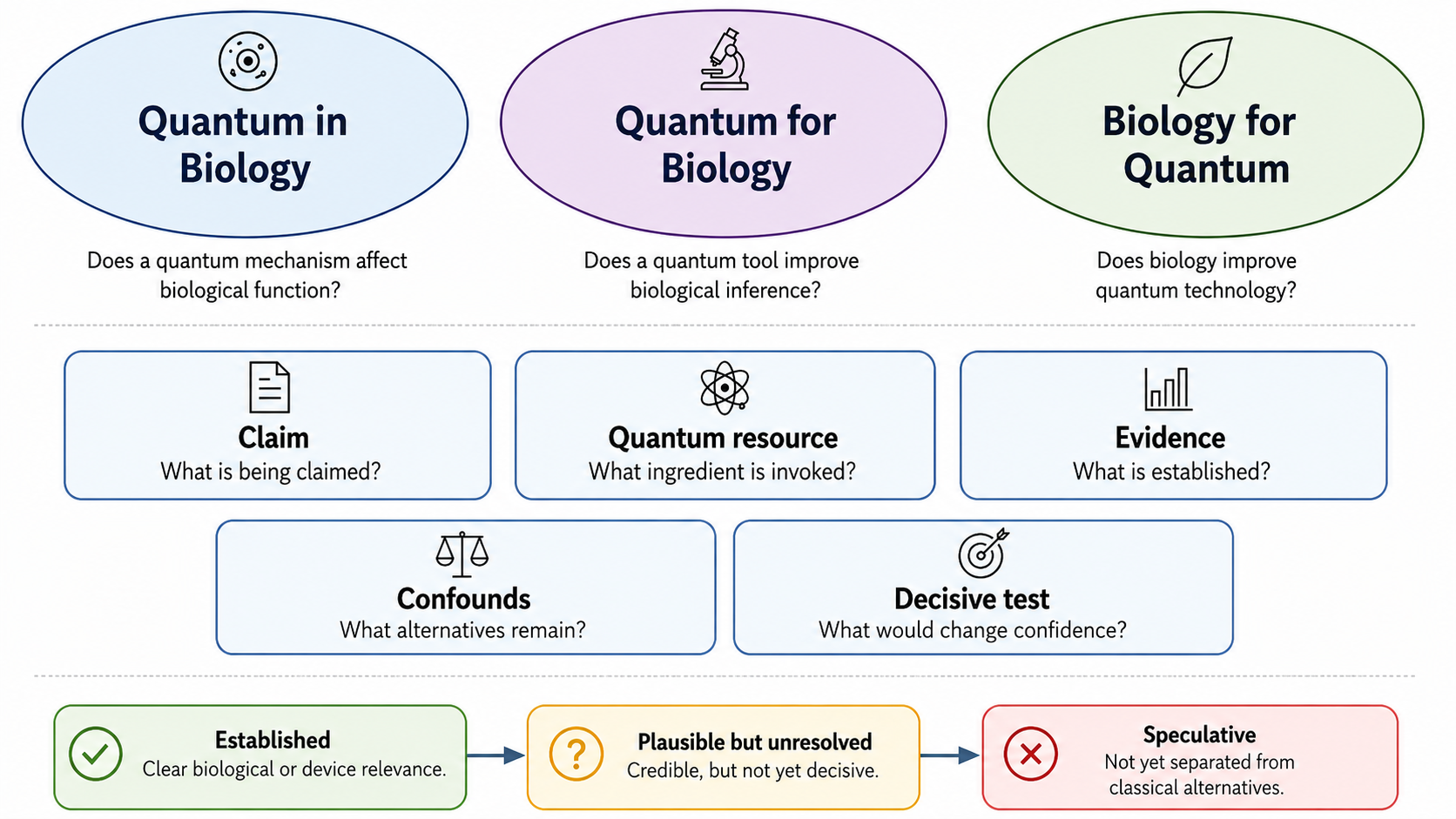}
    \caption{Conceptual framework used throughout the review. Each topic is mapped across three directions: quantum in biology, quantum for biology, and biology for quantum. The framework evaluates the central claim, the quantum resource invoked, the supporting evidence, the main classical or engineering confounds, and the tests or benchmarks most likely to change confidence. These judgments are summarized through an evidence ladder ranging from established contributions to plausible but unresolved cases and more speculative proposals.}
    \label{fig:review_framework_evidence_ladder}
\end{figure*}

This review is organized as a navigational map across three directions: quantum in biology, quantum for biology, and biology for quantum. Its purpose is not to provide the longest possible account of each subfield, but to make the status of different topics comparable. For each topic, the review asks five questions: what is the central claim, what quantum resource is being invoked, what do the strongest experiments and models establish, which classical alternatives or engineering confounds remain competitive, and what decisive tests or benchmarks would most efficiently change confidence. This structure is summarized schematically in Figure~\ref{fig:review_framework_evidence_ladder}, which shows how the three review directions are evaluated using the same evidence-ladder criteria.

This is a structured narrative review rather than a formal systematic review or meta-analysis. Reference selection therefore prioritizes papers that constrain mechanism, establish benchmark standards, provide strong counterarguments, or document developments that materially change plausibility or engineering readiness. 

For clarity, we use several quantum terms in a restricted mechanistic sense. Coherence refers to a maintained phase relationship between quantum amplitudes, not simply to coordinated biological behavior. Entanglement refers to nonclassical correlations between subsystems that cannot be reduced to ordinary shared history or classical statistical dependence. Tunneling refers to barrier penetration by a particle, such as an electron, proton, or hydrogen nucleus, through a classically forbidden region. Vibronic mixing refers to coupling between electronic and vibrational degrees of freedom, which can shape spectra and dynamics without necessarily implying long-lived electronic coherence. Spin-dependent dynamics refers to reaction or transport processes whose yields or pathways depend on spin states, such as singlet--triplet interconversion in radical pairs.

A recurring source of confusion is that ``quantum'' can mean different things in different communities. Quantum mechanics is unavoidable for molecular energetics, bonding, and spectroscopy, but that does not imply that long-lived coherence, entanglement, or nonclassical correlations are functionally required. In this review, a topic is treated as quantum in biology only when the claim depends on a quantum ingredient that would be mischaracterized by a purely classical description at the level of mechanism or predicted observables. A topic is treated as quantum for biology when a quantum-enabled technique is evaluated as a practical biological tool against strong classical baselines. A topic is treated as biology for quantum when a biological principle, structure, or fabrication strategy measurably improves or enables a quantum device, material platform, or algorithm.

Throughout the paper, it is useful to keep an informal evidence ladder which should be understood as a qualitative guide rather than a numerical ranking. At the strongest level are established quantum-chemical or quantum-technological contributions with measurable biological or engineering consequences. An intermediate level contains mechanisms that are physically plausible and experimentally suggestive, but not yet functionally decisive under physiological or operational conditions. A more speculative level contains proposals that currently lack discriminating evidence or remain weakly separated from classical alternatives. Many topics contain elements from more than one level, so the purpose of the ladder is to prevent category errors: a topic may be quantum mechanical at the microscopic level without yet showing biological necessity, and a quantum technology may be physically sophisticated without yet improving biological inference in practice.

\section{Quantum in Biology}

\begin{figure*}[t]
    \centering
    \includegraphics[width=\textwidth]{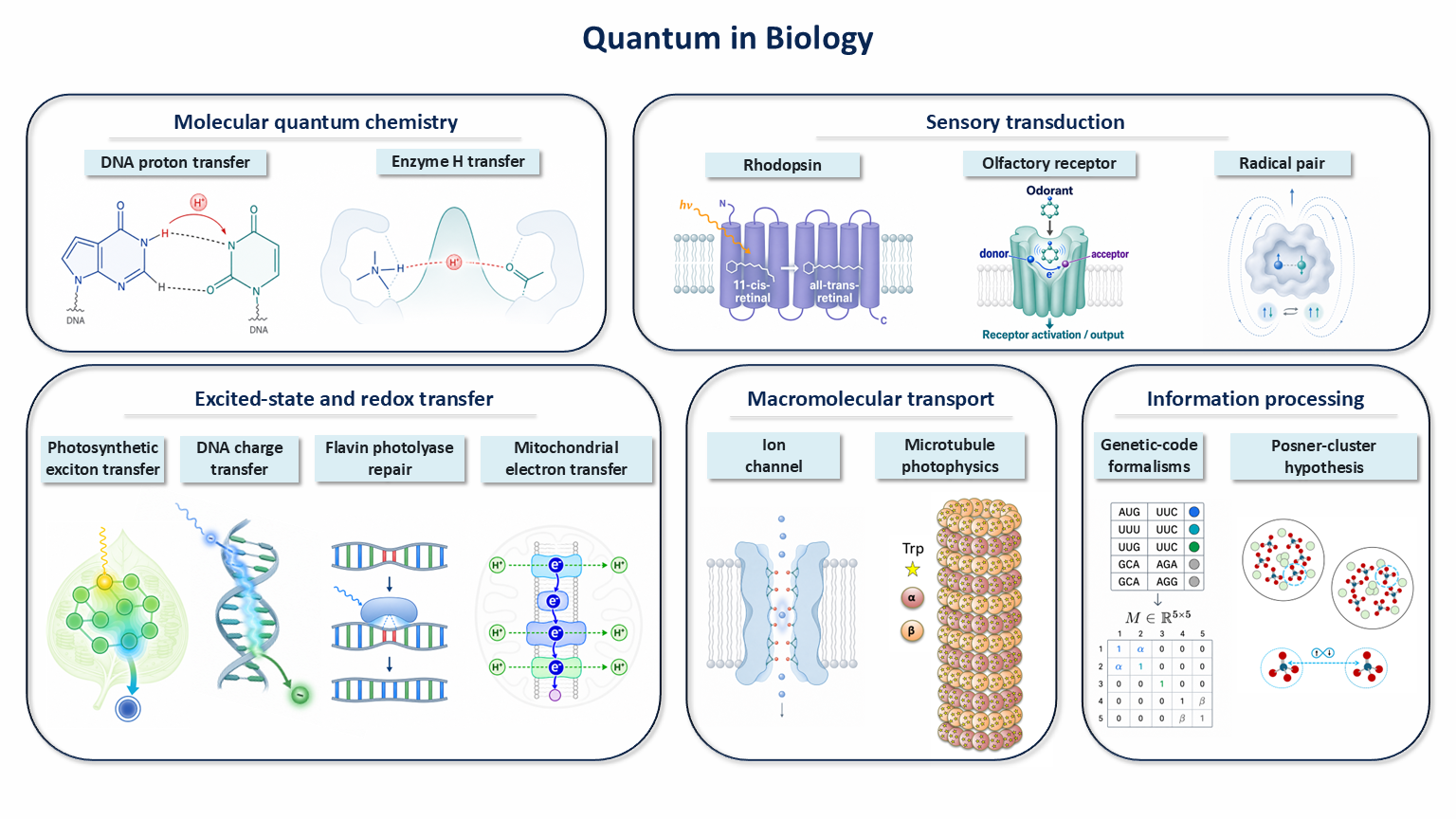}
   \caption{Overview of selected topic areas in which quantum mechanical concepts are established, proposed, or debated in biological systems. The panels schematically illustrate molecular quantum chemistry, including DNA proton transfer and enzymatic hydrogen transfer; sensory transduction, including retinal photoisomerization in rhodopsin, olfactory receptor-level vibrational or electron-transfer hypotheses, and radical-pair magnetoreception; excited-state and redox processes, including photosynthetic exciton transfer in a leaf/chloroplast context, DNA charge transport, flavin photolyase repair, and mitochondrial electron transfer in a mitochondrial context; macromolecular transport, including ion-channel selectivity and microtubule photophysics; and proposed information-processing formalisms, including matrix- or operator-based representations of genetic-code structure and the Posner-cluster hypothesis. The figure is illustrative and does not imply that all topics have the same evidential status or biological relevance.}
    \label{fig:quantum_in_biology_overview}
\end{figure*}

Quantum in biology asks whether specific quantum mechanisms contribute to biological function in ways that would be mischaracterized by purely classical or semiclassical accounts. The goal of this section is to remain concrete about the nature of the claim, the conditions under which it is proposed to operate, the strongest competing explanations, and the decisive tests required to evaluate it \cite{Lambert2013,Marais2018}. Figure~\ref{fig:quantum_in_biology_overview} summarizes the principal topic areas discussed in this section, spanning molecular quantum chemistry, sensory transduction, excited-state and charge-transfer processes, spin-dependent mechanisms, transport hypotheses, and information-processing proposals. Across these cases, the evidential burden increases markedly as one moves from microscopic plausibility to biological necessity: it is easier to support a quantum correction to a local observable than a system-level claim that a quantum resource is required for function in vivo. The guiding question throughout is therefore not merely whether a quantum description exists, but whether it is necessary to explain the relevant biological observable.

\subsection{Molecular quantum chemistry and transformation}

\subsubsection{DNA replication and proton transfer}

A focused quantum claim in DNA replication concerns proton transfer along hydrogen bonds in base pairs and the possibility of transient tautomerization that alters pairing preferences during polymerase selection \cite{loewdin1963proton,pusuluk2018quantum}. In physical terms, the relevant coordinate is the position of a proton shared between donor and acceptor atoms in a base-pair hydrogen bond. If that proton transiently occupies the alternative well, the base can enter a rare tautomeric or ionized form whose hydrogen-bonding pattern resembles a different pairing state. During polymerase copying, such a state could in principle increase the probability that the enzyme accepts an incorrect nucleotide.

The candidate quantum resource is not limited to nuclear tunneling through an effective barrier, but may also involve proton delocalization along a partially covalent hydrogen bond, with the balance between these descriptions depending on local geometry, electrostatics, hydration, and base stacking \cite{pusuluk2018quantum}. Support is strongest at the level of quantum chemistry, which indicates that hydrogen bonds in DNA base pairs are not purely electrostatic interactions but can exhibit charge-transfer and resonance-assisted covalent character, arising from overlap between the donor X--H antibonding orbital and the acceptor lone-pair orbital \cite{Weinhold2014,FonsecaGuerra1999}. In that setting, proton motion can be more appropriately viewed in some cases as transient delocalization within a coupled donor--acceptor coordinate rather than only as a classical over-barrier process corrected by tunneling. Biochemically realistic open-system models have also shown that the partial covalency of hydrogen bonds can support nonclassical correlations during proton transfer in related biomolecular settings \cite{Pusuluk_2018_PRSA,Pusuluk_2019_PRSA}, but comparable models directly linking proton dynamics in DNA replication to mutation statistics remain much less developed.

The main biological confound is that DNA replication is not controlled by a single proton-transfer event. Polymerases first bind and select a nucleotide, then undergo conformational checks, catalyze bond formation, and often proofread or reject errors after insertion. Therefore, even if proton dynamics changes a microscopic tautomerization rate, that effect may be masked or corrected by later enzyme steps and may not determine the final mutation rate \cite{freudenthal2013observing,jamsen2022watching}. In addition, rare Watson--Crick-like mispairs can arise from low-populated transient tautomeric or ionized states, which complicates any simple attribution of replication error rates to tunneling alone \cite{wang2011structural,kimsey2015visualizing}. A useful next step would be polymerase-resolved assays combining isotope substitution and temperature scaling with constraints on tautomer lifetimes and nucleotide misincorporation probabilities, while controlling for classical binding and conformational effects and separating chemical selection from post-insertion proofreading.

\subsubsection{Enzyme catalysis}

Enzymatic hydrogen transfer provides mature cases in which tunneling contributions are supported by multiple observables. In a classical transition-state picture, the transferring hydrogen isotope crosses an activation barrier mainly by thermal activation, with isotope effects arising from differences in mass and vibrational zero-point energy. This picture can explain some isotope dependence, but it is often insufficient when the measured kinetic isotope effects, their temperature dependence, and the activation parameters cannot be reproduced by a single classical barrier under independently constrained structural conditions \cite{klinman2013hydrogen}. The magnitude of the effect is usually assessed through kinetic isotope effects, such as $k_{\mathrm{H}}/k_{\mathrm{D}}$ or $k_{\mathrm{H}}/k_{\mathrm{T}}$, where D and T denote deuterium and tritium, respectively, so these ratios compare reaction rates after substituting hydrogen with heavier isotopes, and through how these ratios change with temperature. Large or weakly temperature-dependent isotope effects, especially when combined with pressure, viscosity, or mutational perturbations, are therefore used as practical signatures that barrier penetration contributes to the measured rate \cite{klinman2013hydrogen}.

The candidate quantum resource is tunneling of the transferring nucleus, often coupled to enzyme molecular motions that modulate the donor--acceptor distance, the reaction coordinate, and the effective barrier width. These motions are sometimes described as gating or promoting vibrations \cite{klinman2013hydrogen}. While some earlier theoretical treatments based on relatively static barrier descriptions or limited conformational sampling argued that tunneling need not be catalytically decisive, more recent dynamically modulated and open-system models suggest that enzyme-driven conformational fluctuations can enhance or sustain nonclassical proton-transfer contributions under biologically relevant noisy conditions \cite{Jevtic2017,Pusuluk_2018_PRSA}.

Evidence is strongest when isotope effects, temperature trends, pressure or viscosity perturbations, and series of active-site or second-shell substitutions align with constrained theoretical models that do not rely on unconstrained fit parameters \cite{klinman2013hydrogen,korchagina2025directed}. A key confound is that, in multi-step catalytic cycles, the rate-limiting step can shift, producing apparent tunneling signatures even when the chemical step itself is not cleanly isolated; another confound is that enzyme molecular motions can change electrostatics and reorganization energies, complicating attribution. Decisive studies therefore combine targeted amino-acid substitutions in the active site or second shell, substrate or solvent isotope substitution, and dynamical modeling constrained by independent structural data, and, in some cases, directed-evolution series that tune tunneling signatures systematically while tracking changes in the rate-limiting step \cite{korchagina2025directed}.

\subsubsection{Proton-coupled electron transfer and proton transport motifs}

Across enzymes and bioenergetics, proton-coupled electron transfer provides a unifying mechanistic motif that is quantum mechanical at the level of the transferring particles but not necessarily nonclassical in the sense of long-lived coherence. Proton-coupled electron transfer means that movement of an electron is coupled to movement or repositioning of a proton, either in the same elementary step or in a tightly linked sequence of steps. This coupling matters because electron transfer changes charge distribution and acidity, while proton motion changes electrostatics and can therefore tune the electron-transfer driving force and barrier. The candidate quantum ingredient is tunneling and nonadiabatic coupling in coupled electron and proton coordinates, often modulated by hydrogen-bond geometry, electrostatics, and reorganization in the surrounding environment \cite{Mayer2004ARPC,HammesSchiffer2010ChemRev,HammesSchiffer2015JACS}. Evidence is strong at the level of physical chemistry theory and targeted model systems, and many biological redox enzymes and respiratory complexes can be framed naturally in a proton-coupled electron-transfer language. The main confound for strong quantum-biology claims is again attribution: environmental fluctuations can dominate kinetics, and many observables are reproduced by semiclassical or mixed quantum--classical rate theories with effective parameters \cite{HammesSchiffer2010ChemRev,HammesSchiffer2015JACS}. Decisive studies in enzymes and respiratory complexes should therefore isolate coupled electron--proton steps within full catalytic cycles and test predicted scaling with isotope substitution, pH dependence, and electrostatic perturbations while holding the redox landscape as constant as possible.

\subsection{Sensory transduction}
\subsubsection{Vision}

Early vertebrate vision is often described as operating near the quantum limit because rod photoreceptors can respond at the level of individual photons, implying that the first stage of visual transduction is constrained by quantum detection statistics and thermal noise suppression in the dark state \cite{hecht1942energy,rieke1998single,reingruber2013detection,sia2014quantum,moazed2023quantum}. At the molecular level, retinal in rhodopsin undergoes ultrafast cis-to-trans isomerization after photon absorption through a conical-intersection landscape shaped by the protein pocket \cite{hubbard1958action,kropf1970photoisomerization,huntress2013toward}. The mechanistic question is whether transient vibronic coherence meaningfully influences photochemical yield, branching ratios, or timing beyond what is expected from the energy landscape and classical-like relaxation on that landscape \cite{marsili2020quantum,tscherbul2015quantum,ben1998quantum}. Evidence is strong for ultrafast dynamics and environmental tuning, but the necessity of coherence for function remains conditional because vibrational wavepackets and inhomogeneous ensembles can also generate oscillatory signals, and because natural-light excitation changes the initial preparation of coherence \cite{sen2022insight,talotta2022describing,pereira2023quantum}. A decisive route would be to perturb dephasing or vibronic coupling while preserving the underlying energetics and structure as much as possible, and then test for predicted changes in quantum yield, isomerization-time distributions, or downstream amplification statistics in intact systems.

A complementary line of work has treated retinal photoisomerization within the quantum resource theory of thermodynamics, using thermomajorization and related state-conversion tools to derive operational bounds on switching efficiency rather than to resolve the microscopic dynamics in detail \cite{yunger2020fundamental, spaventa2022capacity, burkhard2024boosting, tiwary2025quantum}. To date, this remains one of the clearest biomolecular settings in quantum biology where such methods have yielded experimentally interpretable and potentially falsifiable predictions. 

\subsubsection{Olfaction and vibrationally assisted receptor activation}

The vibrational theory of olfaction proposes that receptors may discriminate odorants through inelastic electron tunneling, in which odorant vibrational modes are hypothesized to modulate an electron-transfer step, rather than through shape complementarity alone \cite{bittner2012quantum,solov2012vibrationally,tirandaz2017validity}. The proposed quantum ingredient is therefore vibrationally assisted electron transfer in a dissipative receptor environment, where the putative inelastic electron-transfer pathway depends on energy matching to internal vibrational modes and on the coupling between donor and acceptor states in the receptor context \cite{solov2012vibrationally,tirandaz2017validity}. In the related ``swipe-card'' picture, the proposed vibrational mechanism complements rather than replaces conventional lock-and-key recognition: molecular shape and noncovalent interactions allow the odorant to bind and adopt a suitable orientation, while vibrational modes may provide an additional layer of discrimination \cite{Brookes2012SwipeCard,Horsfield2017Olfaction}.

Evidence in favor remains largely theoretical, but some of the stronger open-system analyses suggest that environmental coupling and odorant-mode dissipation can play a constructive role in enhancing both transfer selectivity and vibrational frequency resolution, implying that isolated receptor assays may not fully test the proposed mechanism under physiologically relevant conditions \cite{checinska2015dissipation}. Evidence against emphasizes biochemical feasibility and the lack of receptor-level measurements that uniquely require inelastic tunneling rather than conventional binding and activation mechanisms \cite{block2015implausibility,hoehn2018status}. A major confound is that isotopic substitution can also alter noncovalent interaction energies, including binding contributions, so behavioral or perceptual differences are not uniquely diagnostic of an inelastic tunneling mechanism \cite{pokora2023noncovalent,krzeminska2015binding}.

Decisive progress would therefore come from receptor-resolved assays that isolate responses at the level of individual olfactory receptors while using carefully matched isotopologues designed to minimize nonvibrational differences in binding affinity and binding pose. Stronger evidence would also require direct detection of putative mechanistic signatures, such as transient charge-transfer intermediates or current-like signals associated with inelastic tunneling, together with perturbations that selectively disrupt the proposed tunneling pathway while leaving overall ligand binding largely unchanged. 

Although originally developed in the context of olfaction, related vibrational hypotheses have also been extended to other ligand--receptor systems. Studies of serotonin, histamine, and adenosine receptors have reported characteristic vibrational patterns associated with ligand potency or functional classification, suggesting that vibrational spectra may provide complementary molecular descriptors beyond structure alone \cite{Hoehn2015Neuroreceptor,Hoehn2017GPCRVibration,Oh2012HistamineVibration,Chee2013AdenosineVibration,Chee2015AdenosineVibrations}. More recently, a possible extension to ligand-gated ion channels has also been proposed, although this remains a hypothesis requiring direct receptor-level and mechanistic validation \cite{Pagan2024LigandReceptorVibrations}.

\subsubsection{Magnetoreception}

The radical-pair mechanism is a leading hypothesis for magnetoreception. Photoinduced electron transfer generates radical pairs whose singlet--triplet interconversion is sensitive to weak magnetic fields through hyperfine and Zeeman interactions and, in more detailed treatments, can also be influenced by exchange and dipolar couplings, thereby allowing directional sensitivity in principle \cite{Ritz2000,hiscock2016quantum,fay2020quantum,chen2024identifying}. The candidate quantum resource is spin correlation and coherent spin dynamics in an open-system environment, where decoherence sets the sensitivity window but does not necessarily eliminate directional information \cite{bandyopadhyay2012quantum,tiersch2012decoherence,adams2018open}.

The most informative evidence comes from convergence rather than from any single decisive experiment: physically constrained spin-dynamics models, proxy spin-chemistry experiments, controlled simulations, and chemically realistic parameter estimates all support the feasibility of radical-pair magnetic sensitivity under some conditions \cite{hiscock2016quantum,cai2013chemical,fay2020quantum,chen2024identifying,kominis2025approaching}. Major uncertainties concern the identity and lifetime of the relevant radical pairs under natural illumination, how those radicals are embedded in the relevant protein and cellular context, the role of competing relaxation channels, and the transduction pathway from reaction-yield changes to neural signals and behavior \cite{xie2022searching}.

Isotopic substitution provides an additional mechanistic probe of radical-pair magnetosensitivity because changes in nuclear spin and magnetic moment modify the hyperfine interactions that govern singlet--triplet dynamics. Spin-dynamics simulations of cryptochrome-based radical pairs predict that substitutions such as $^1$H$\rightarrow{}^2$H, $^{12}$C$\rightarrow{}^{13}$C, and $^{14}$N$\rightarrow{}^{15}$N can alter magnetic sensitivity, suggesting targeted isotope substitution as a possible discriminating test of radical-pair models \cite{Pazera2023IsotopeCryptochrome,Galvan2024IsotopeCryptochrome}. 

Alternative mechanisms such as magnetite-based sensing may operate in some taxa and can confound broad cross-species claims \cite{kirschvink1982birds,xie2022searching}. Recent hypomagnetic-field experiments in \textit{Xenopus} embryos further show that weak magnetic-field effects can be measurable in biological development under controlled conditions, although the mechanism remains unresolved and should not be assumed to be radical-pair chemistry without additional tests \cite{Lodesani2024WeakMagnetic}. The most discriminating future work should therefore combine genetics, protein biochemistry, spectroscopy, and controlled electromagnetic perturbations, but these tests should be interpreted as a convergent program rather than as individually decisive. Confidence would increase most strongly if predicted light dependence, spectral dependence, angle dependence, isotope effects, and genetic or biochemical perturbations were reproduced across laboratories within the same candidate molecular pathway.

\subsection{Excited-state, redox, and spin transfer}
\subsubsection{Photosynthesis}

Ultrafast spectroscopy has revealed oscillatory signals in some photosynthetic complexes that can be described using coherent or vibronically mixed excitonic dynamics \cite{Engel2007,Panitchayangkoon2010,Collini2010}. The candidate quantum resource is delocalization and coherence in excitonic manifolds coupled to vibrational modes, with the possibility that intermediate system--bath coupling can support efficient transport and robustness, often discussed in the context of noise-assisted transport \cite{ishizaki2009theoretical,levi2015quantum,chen2015using,plenio2008dephasing}.

Evidence is strong that coherence-like signatures can be measured in isolated complexes and that open-system models can reproduce many observables across temperature and disorder regimes. The central confound is that vibrational motion, ensemble effects, and static disorder can mimic electronic-coherence signatures, and functional advantage in vivo is harder to establish because physiological conditions include regulation, photoprotection, and dynamic reconfiguration \cite{runeson2022explaining,keren2018photosynthetic}.

A decisive route would be to perturb specific vibronic resonances or dephasing channels while keeping pigment geometry and site energetics as fixed as possible, and then test whether the resulting changes track only spectroscopic oscillations or also measurable transport efficiency under near-native conditions. In practice, fully separating purely spectroscopic coherence from transport-relevant delocalization remains experimentally challenging, so this should be treated as a demanding benchmark rather than as an already established capability.

\subsubsection{DNA charge transport and redox signaling}

At the interface of chemistry and biology, another potentially quantum-relevant theme is charge transport through DNA and along protein-mediated pathways, which can couple electron transfer to DNA damage detection and repair-protein signaling \cite{Merino2008DNACT,Sontz2012ACCR,genereux2010mechanisms,arnold2016dnact}. Here, a lesion means a damaged base, mismatch, or photochemical product that distorts the DNA stack, while redox signaling means communication through changes in oxidation state, often involving electron transfer between DNA-bound repair proteins. The most defensible quantum ingredient is charge migration through the $\pi$-stacked base manifold, which at short range can proceed through superexchange-like tunneling and over longer distances is often described in terms of thermally assisted hopping between favorable sites, with strong sensitivity to base-stacking integrity \cite{genereux2010mechanisms,arnold2016dnact}.

The strongest evidence base is physicochemical: experiments show that DNA-mediated charge transport is highly sensitive to perturbations that disrupt base stacking, and that redox-active repair proteins can participate in pathways consistent with DNA-mediated redox communication under controlled conditions \cite{Merino2008DNACT,Sontz2012ACCR,arnold2016dnact}. A key confound is that extending these results to lesion search and repair in cells remains difficult, because repair outcomes also depend on protein recruitment, diffusive search, chromatin organization, and lesion-recognition steps in the broader cellular context, so DNA-mediated transport need not be rate limiting even if it contributes \cite{Merino2008DNACT,Sontz2012ACCR,arnold2016dnact}. A useful discriminating approach would therefore be quantitative perturbation experiments in nucleosome-containing or chromatin-like substrates and in living cells. In such tests, redox-active mutants would mean repair-protein variants in which the relevant electron-transfer cofactor or redox pathway is altered, while controlled lesion densities would mean DNA substrates or cellular systems in which the number and spacing of damage sites are known. The goal would be to test whether charge-transport-mediated recruitment or signaling explains the data better than conventional diffusion-and-binding models.

\subsubsection{DNA repair photochemistry and ultrafast electron transfer}

A canonical example in which quantum photochemistry directly underpins biological function is DNA photolyase, where absorption by a flavin cofactor initiates ultrafast electron-transfer steps that drive repair of ultraviolet-induced lesions \cite{Sancar2003Photolyase,Thiagarajan2011PNAS}. In this case, the most secure quantum-mechanical ingredient is the electronically excited-state photochemistry and subsequent electron-transfer dynamics of the cofactor--substrate system in its protein environment \cite{Sancar2003Photolyase,Thiagarajan2011PNAS}. The evidence base is strong for the overall repair mechanism, the identity of the key cofactors, and the timescales and directionality of the main transfer steps in well-characterized systems \cite{Sancar2003Photolyase,Thiagarajan2011PNAS}. Stronger claims about functionally important coherence, however, remain harder to establish, because much of the observed behavior can be described within established electron-transfer frameworks in fluctuating protein environments \cite{Narth2015ACR}. Decisive tests would therefore require perturbations that selectively alter excited-state dephasing or related dynamical couplings while leaving the main driving forces and structure as unchanged as possible, and then showing corresponding changes in repair yield or pathway branching under near-physiological excitation conditions.

\subsubsection{Cellular respiration}

Mitochondrial respiration transfers electrons from metabolic substrates through redox cofactors in the respiratory complexes of the inner mitochondrial membrane, coupling stepwise electron transfer to proton translocation and ATP synthesis. The most plausible quantum contributions in this context are tunneling corrections to electron- and proton-transfer steps, because these processes involve donor--acceptor distances and barriers for which quantum effects can become rate-relevant, particularly when coupled to protein reorganization, gating motions, and proton-coupled electron-transfer chemistry \cite{moser2006electron,verkhovskaya2008real,hayashi2010electron,hayashi2011quantum,hayashi2011electron,martin2017electron,bennett2021energy,abdallat2024mathematical,Mayer2004ARPC,HammesSchiffer2010ChemRev,HammesSchiffer2015JACS}.

More speculative proposals have framed mitochondria as candidate quantum-biological systems, but the secure claim is narrower: local respiratory-chain chemistry can require quantum-mechanical electron and proton transfer, whereas the proton-motive force is a classical electrochemical gradient rather than a collective quantum state \cite{nunn2016quantum}. At larger scales, cristae ultrastructure, respiratory-supercomplex organization, fusion--fission dynamics, and inter-mitochondrial junctions shape membrane potential, diffusion, reactive oxygen species (ROS) production, and respiratory efficiency through mechanisms that are presently best interpreted within classical bioenergetics and cell biology \cite{Picard2015CristaeCoordination,Picard2022Signal,Wallace2005Mito}. Recent NADH-anisotropy measurements support strong mitochondrial structural directionality in living cells, but whether this alignment has quantum-relevant electrochemical or photonic consequences is not yet established \cite{Smith2024NADHAnisotropyCristae}. Mitochondria are also plausible contributors to ultra-weak photon emission through ROS-driven excited-state chemistry, and recent reports of non-chemical signaling between isolated mitochondria are intriguing, but not yet decisive evidence for a functional quantum-optical communication channel \cite{VanWijk2020MitochondrialUPE,Mould2023NonChemicalMitochondria}. Decisive studies should therefore isolate perturbations that alter tunneling likelihood while minimally changing classical driving forces, redox landscape, and membrane potential, and then link those microscopic changes to biochemical or cellular outputs.

\subsubsection{Ultra-weak photon emission}

Ultra-weak photon emission (UPE), often called biophoton emission, is the spontaneous emission of very low-intensity light from biological systems. It is generally understood as an endogenous photophysical signal linked to oxidative metabolism, reactive oxygen species, and electronically excited molecular species generated during lipid, protein, and nucleic-acid oxidation \cite{PospisilPrasadRac2019ExcitedSpecies,Mould2024UPEReview}. Recent studies support UPE as a possible non-invasive marker of biological state, including cancer-related metabolic changes and dynamic physiological responses to ischemia--reperfusion \cite{Murugan2020CancerUPE,Belksma2026IschemiaReperfusionUPE}. Radiation-bystander studies further suggest that UPE or radiation-induced biophoton signals can participate in downstream cellular responses, including exosome-mediated effects, but these observations are best interpreted at present as photon-triggered photochemical or biochemical signaling rather than evidence for nonclassical light \cite{Le2017BiophotonExosomes,Tong2024BiophotonSignaling}.

For biological interpretation, UPE is best viewed as an integrated redox-optical signal rather than a source-specific reporter. The detected emission can combine contributions from mitochondrial respiration, lipid peroxidation, singlet-oxygen chemistry, inflammatory redox activity, tissue optical filtering, and detector response, and is therefore most informative when interpreted alongside independent metabolic, electrophysiological, and ROS measures \cite{PospisilPrasadRac2019ExcitedSpecies,Mould2024UPEReview,VanWijk2020MitochondrialUPE}.

For quantum biology, the key distinction is between UPE as an established photophysical observable and stronger claims about functional quantum resources. Although UPE involves quantum transitions from excited molecular states to lower-energy states, this does not by itself imply long-lived optical coherence, entanglement, nonclassical light, or biologically meaningful photon-mediated communication. Critical analyses of photocount statistics conclude that evidence for coherence or nonclassicality in biological UPE remains limited \cite{Cifra2015BiophotonsCoherence}. Decisive progress will require calibrated spectral, temporal, and photon-statistical measurements under controlled physiological or redox perturbations, with independent metabolic, electrophysiological, ROS, and tissue-optical validation.

\subsubsection{Chiral induced spin selectivity in biological electron transfer}

Chiral induced spin selectivity, often abbreviated CISS, describes the observation that electron transmission through a chiral structure can depend on the electron spin orientation. In biological terms, the relevant point is that many biomolecules, including DNA, proteins, and peptide assemblies, are chiral, so their geometry could in principle couple molecular handedness to spin-dependent electron transport or spin-sensitive chemistry \cite{bloom2024chemrev,aiello2022chirality,Naaman2022CISSBiology}. The candidate quantum ingredient is spin-dependent transport associated with chiral molecular structure, typically discussed in the context of spin--orbit-coupled electron transmission, interfacial charge transport, and spin-dependent recombination processes \cite{bloom2024chemrev,aiello2022chirality,Naaman2019ChiralMolecules}. Biological relevance is suggested by experiments on spin filtering through DNA duplexes and by broader CISS studies in biomolecular and bioinspired systems \cite{Zwang2016DNASpinFiltering,Naaman2022CISSBiology}. 

Evidence for CISS itself is strong in physicochemical junction and interface experiments, whereas its direct biological role remains less certain because many demonstrations rely on engineered contacts, surface immobilization, applied electrodes, or device geometries that do not directly map onto solution-phase biological electron transfer \cite{bloom2024chemrev,aiello2022chirality,Naaman2022CISSBiology}. Important confounds include electrode and interface effects, uncontrolled contact geometry, and classical explanations based on distance, electrostatics, or protein dynamics. Decisive tests should therefore focus on minimally invasive solution-phase or near-physiological measurements in which chirality is perturbed while classical parameters are carefully controlled, together with independent detection of spin-polarization signatures and clear linkage to a biological or biochemical outcome.

\subsubsection{Ferritin protein nanoparticles, electron tunneling, and nanomagnetism}

Ferritin provides a distinct protein-nanoparticle case at the boundary between biomolecular electron transport, nanomagnetism, iron redox chemistry, and biological interpretation. Ferritin is a chiral protein cage containing an iron-oxide core, and biochemical reduction studies indicate that electron transfer to the ferritin core and iron release can occur without requiring reductant entry into the ferritin interior \cite{Watt1988FerritinRedox}. Ferritin-based junction and multilayer experiments have further reported long-range electron tunneling, iron-loading-dependent transport regimes, and Coulomb-blockade-like behavior under device conditions \cite{Kumar2016FerritinTunneling,Bera2019FerritinMultilayers,Rourk2021FerritinDMFS,DiezPerez2023FerritinTunneling}.

Related work on chiral-molecule-coated superparamagnetic iron oxide nanoparticles has shown that asymmetric adsorption of chiral molecules can stabilize single-domain ferromagnetism at approximately 10 nm, with the magnetic orientation depending on molecular handedness \cite{Koplovitz2019ChiralSPION}; a later study reported related magnetic-monopole-like behavior in superparamagnetic nanoparticles coated with chiral molecules \cite{Zhu2024ChiralSPION}. These findings do not demonstrate the same effect in ferritin in vivo, but they make ferritin, because it combines a chiral protein cage with an iron-oxide core, a plausible biomolecular system in which electron transport, spin selectivity, and nanomagnetism could intersect.

Ferritin is therefore relevant to quantum-biology discussions of electron tunneling and biomolecular nanomagnetism, but it should not yet be treated as evidence for an established in vivo signaling mechanism. Recent review-level proposals have suggested that single-particle or array-level ferritin electrical and magnetic properties could influence cellular bioelectric or biomagnetic processes \cite{Rourk2025FerritinBioelectricity}. At present, however, the biological claims remain hypothesis-level because device geometries, iron loading, hydration, aggregation state, redox environment, and cellular localization can all change the relevant transport and magnetic observables. Decisive tests would require in situ or near-native measurements that vary ferritin iron loading, redox state, magnetic field, and chiral or spin-sensitive readouts while linking the result to a defined biological endpoint.

\subsection{Macromolecular transport and collective effects}

\subsubsection{Ion channels and quantum transport hypotheses}

Ion channels and proton channels are often discussed in quantum-biology debates because their selectivity filters and hydrogen-bond networks involve nanometer length scales, strong confinement, and rapid transport. The strongest established physics here is conventional quantum chemistry: electronic structure helps determine the energetics of ion dehydration, coordination of permeating ions by filter ligands such as backbone carbonyl oxygens, and proton transfer along hydrogen-bond networks or water-wire pathways \cite{Roux2011IonSelectivity,DeCoursey2003ProtonChannels}. Stronger proposals suggest that a candidate quantum resource, contributing to selectivity or conduction, could be transient coherent transport or resonant delocalization of ions or protons across a small number of filter sites with environmental fluctuations setting the relevant dephasing rates  \cite{VaziriPlenio2010NJP,Ganim2011NJP}. The main confounds are that classical stochastic models with detailed energetics and strong coupling to thermal fluctuations can reproduce many conductance and selectivity features, and that experimental observables that uniquely require coherence remain scarce \cite{Roux2011IonSelectivity}. A useful discriminating strategy would therefore be channel-specific perturbations that selectively alter predicted coherence times or resonance conditions without broadly changing classical energetics, together with measurements designed to distinguish coherent signatures from classical hopping in the same structural context.

\subsubsection{Microtubules}

Microtubules have long attracted proposals ranging from optical and electrodynamic roles to excitonic transport in aromatic networks, including discussions of collective radiative effects, network delocalization, and transport behavior that may depart from simple F\"orster hopping under some conditions \cite{jibu1994quantum,Mavromatos1998,mershin2004tubulin,craddock2014feasibility,celardo2019existence,gassab2026optical}. For the purposes of this review, the most defensible question is narrower than the broad historical literature: do aromatic residues in tubulin support measurable excited-state transport or collective optical effects that cannot be reduced to simple incoherent hopping under specified experimental conditions? Early models were important in motivating the field, but did not by themselves establish a physiological mechanism \cite{jibu1994quantum,Mavromatos1998}. More recent work has moved toward testable photophysical mechanisms, including excitonic descriptions of aromatic networks and explicit open-system transport modeling under disorder and bath coupling \cite{shirmovsky2023modeling,patwa2024quantum,babcock2024ultraviolet}.

Recent theoretical and in vitro studies have reported signatures consistent with extended transport, ultraviolet collective effects, and estimated triplet or singlet migration in cytoskeletal polymers \cite{kalra2023electronic,kakati2024triplet,shirmovsky2023modeling,patwa2024quantum,babcock2024ultraviolet}. More recently, the microtubule literature has expanded beyond optical transport with an experimental report that magnetic-isotope substitution changes tubulin polymerization dynamics under controlled in vitro conditions \cite{ZadehHaghighi2026}. Because this result is still recent, its interpretation as evidence consistent with a radical-pair mechanism requires independent replication and mechanistic refinement. Although this finding does not establish physiological quantum functionality in vivo, it offers a more specific experimental handle for probing spin-dependent effects in cytoskeletal biology. 

At present, however, the overall evidence base remains nonconvergent. The strongest claims concern controllable photophysical or spin-sensitive signatures in prepared samples rather than established physiological roles in cells. Important confounds include sample and preparation dependence, competing quenching pathways, uncertain excitation conditions in vivo, and the possibility that structured classical hopping can reproduce some apparently long-range trends \cite{kalra2023electronic,babcock2024ultraviolet}. Progress will require independent replication across laboratories using standardized preparations and explicit sample-quality controls, together with observables that distinguish coherent delocalization from incoherent hopping, such as dephasing-sensitive line shapes, disorder-scaling relations, or perturbations that alter bath coupling while leaving geometry approximately fixed. Broader claims that microtubules support cognitive-scale quantum computation remain speculative and controversial, and are not treated here as evidence for the narrower photophysics question \cite{hagan2002quantum,ja2015anesthetics}. On present evidence, microtubules are best regarded as an interesting but unresolved photophysical candidate rather than an established functional case of quantum biology.

\subsection{Information processing}

\subsubsection{Genetic code and quantum formalisms}

Quantum-inspired approaches to the genetic code should be separated into two different types of claim. The first is mathematical: algebraic, spectral, or quantum-information-inspired formalisms can provide compact ways to describe symmetries, degeneracies, and error-tolerance patterns in codon assignments \cite{fimmel2018genetic,bashford2008spectroscopy}. In this use, ``quantum'' does not necessarily imply physical superposition or coherence in living cells; it is mainly a language for representing structure in a high-dimensional symbolic system. The second, stronger claim is physical: quantum-information concepts are sometimes invoked to discuss the origin of coding, algorithmic search in early evolution, or possible nonclassical constraints on biological information processing \cite{patel2008towards,balazs2016quantum,balazs2006some}. 

The main confound is that many features of the genetic code are already explained by evolutionary optimization, error tolerance, stereochemical constraints, and robustness, and flexible formal models can fit patterns without identifying causal mechanisms \cite{gonzalez2019origin,wills2023origins,wills2019reflexivity}. Stronger evidence would therefore require either a measurable physical quantum signature affecting coding or regulation that cannot be reproduced by classical stochastic models, or a predictive advantage of a quantum formalism under strict model comparison and out-of-sample validation that penalizes flexibility.

\subsubsection{Posner cluster hypothesis}

The Posner cluster hypothesis proposes that calcium phosphate clusters could protect $^{31}$P nuclear spins, thereby enabling unusually long-lived coherence or entanglement that might couple to biochemical signaling \cite{Fisher2015}. Structural and computational work has explored plausible geometries and vibrational fingerprints for such clusters \cite{Swift2018}, while relaxation analyses argue that environmental noise in aqueous settings is likely to shorten coherence times relative to strong versions of the hypothesis and places quantitative constraints on achievable correlation times \cite{PlayerHore2018}. Subsequent work has also raised doubts about the proposed entangling mechanism of phosphorus nuclear spins via pyrophosphate hydrolysis, with realistic spin-dynamics models suggesting that robust entanglement generation and maintenance in such systems is inefficient \cite{korenchan2022limits}. Related studies further suggest that the clusters do not necessarily adopt the high-symmetry configurations originally invoked to protect nuclear spin entanglement, but may instead favor lower-symmetry structures \cite{Agarwal2021}. There is also evidence that coherence-related properties depend strongly on cluster composition and size, with smaller calcium phosphate aggregates sometimes appearing more favorable than larger ones, and with ion substitution, including lithium incorporation, tending to suppress coherence or entanglement-like behavior in modeled systems \cite{Agarwal2023,Adams2025,Patel2025}.

The hypothesis has also attracted interest from a quantum-information perspective, including conceptual comparisons with phosphorus-based spin-qubit architectures in spin-zero environments \cite{Gassab2025,Halpern2019,Adams2025}. A related but distinct recent direction proposes model-agnostic tests of nonclassicality in neural representations, for example Bell-type consistency tests applied to latent representations in autoencoders \cite{Kominis2026BrainBellTest}. This does not directly support the Posner mechanism, but it provides a separate information-theoretic route for asking whether neural data require nonclassical latent-variable structure. Even so, the central uncertainties remain biological and mechanistic: whether Posner-like clusters of the relevant symmetry form and persist in physiological contexts \cite{Agarwal2021}, whether entangled phosphate spin states can be prepared through plausible biochemical pathways \cite{korenchan2022limits}, and whether any nonclassical spin correlations could survive long enough to influence downstream chemistry \cite{PlayerHore2018,Agarwal2023}. The proposal also still lacks a clearly demonstrated quantum-to-biological transduction mechanism, since the suggested links between molecular symmetry, binding, and signaling remain unproven \cite{Fisher2015,Agarwal2021}. Decisive experiments would therefore require direct detection of cluster formation in situ, measurement of spin relaxation and coherence times under relevant conditions, and demonstration of nonclassical spin correlations beyond classical mixtures, ideally with perturbations that selectively affect nuclear-spin environments.

\subsection{Section synthesis}

The quantum-in-biology landscape separates less by biological domain than by evidential maturity. The most secure cases are microscopic quantum-chemical contributions that are tightly constrained by kinetics and mechanism, especially enzymatic hydrogen tunneling and related nonadiabatic proton- or electron-transfer chemistry. An intermediate band contains areas such as radical-pair magnetoreception, photosynthetic exciton and vibronic dynamics, DNA charge transport, DNA repair photochemistry, and respiration-related transfer processes, where the mechanistic frameworks are physically serious but the case for biological necessity remains conditional and perturbation-dependent. A more speculative band contains system-level extrapolations, including strong microtubule-wide physiological claims, ferritin-based bioelectric or biomagnetic signaling hypotheses, Posner-cluster neurobiology, and physical interpretations of quantum-information formalisms in genetic or neural information processing, where decisive tests are still missing or not yet sufficiently discriminating \cite{Lambert2013,Marais2018,kalra2023electronic,Fisher2015}. A compact topic-by-topic summary of this section is provided in Appendix Table~\ref{tab:map_qib}.

\section{Quantum for Biology}

\begin{figure*}[t]
    \centering
    \includegraphics[width=\textwidth]{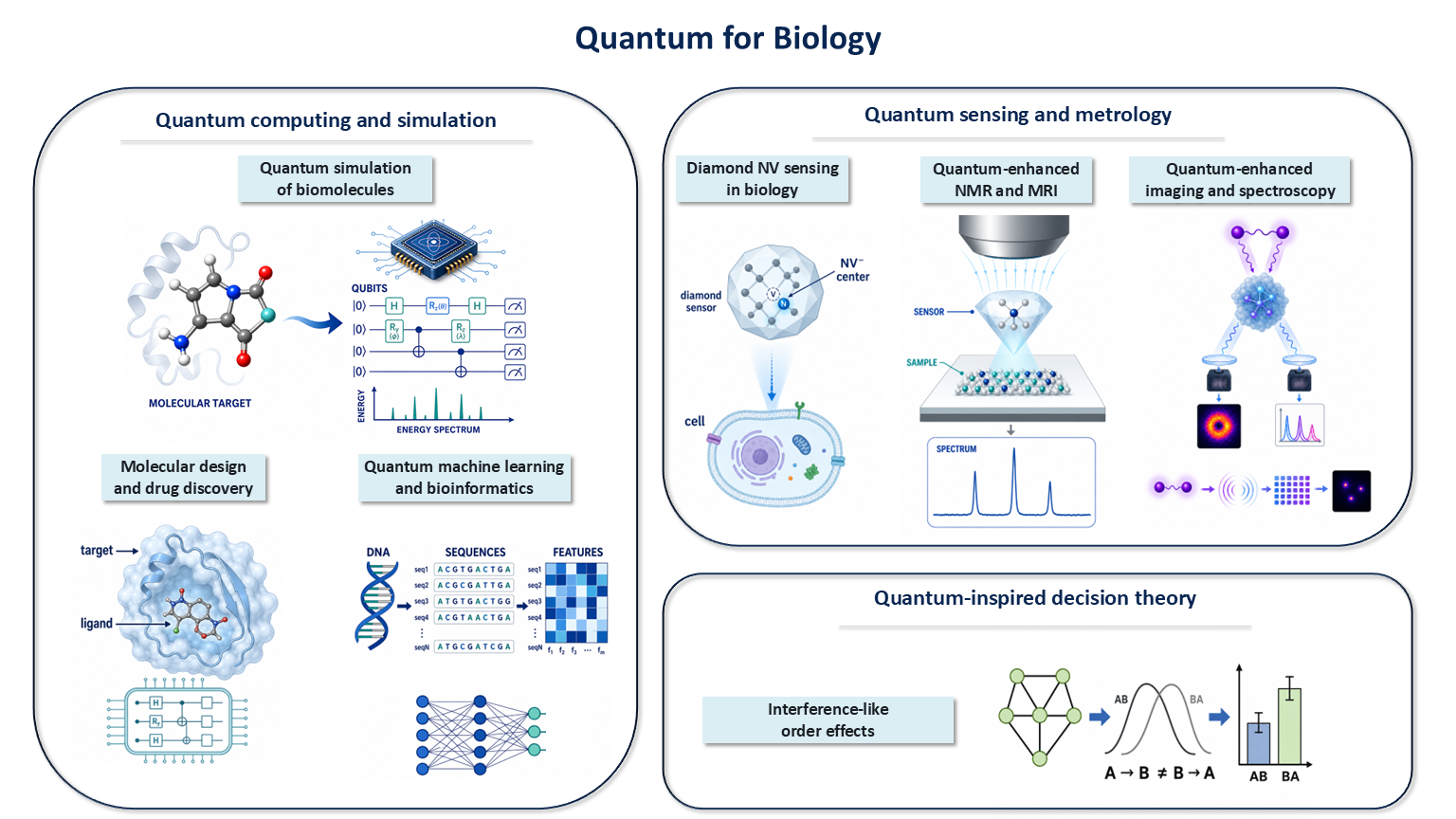}
    \caption{Overview of representative quantum-enabled and quantum-inspired tool classes for biological research. The panels schematically illustrate quantum simulation of biomolecular systems, quantum-assisted molecular design and drug discovery, quantum machine learning for bioinformatics, diamond NV sensing in biological contexts, quantum-enhanced NMR and MRI, quantum-enhanced imaging and spectroscopy, and quantum-inspired decision-theory approaches. The figure summarizes methodological directions discussed in the section and does not imply established quantum advantage for each application.}    
    \label{fig:quantum_for_biology_overview}
\end{figure*}

Quantum for biology uses quantum devices and protocols to improve measurement, imaging, and computation in biological settings. The central question is whether a quantum-enabled approach provides a reproducible advantage under realistic constraints, including calibration, throughput, biological heterogeneity, and comparison against strong classical baselines. Here, ``advantage'' is interpreted operationally: not simply as an improvement in an intermediate technical metric, but as a genuine gain in biological inference or experimental capability once losses, noise, sample variability, and workflow costs are taken into account. Figure~\ref{fig:quantum_for_biology_overview} summarizes the principal directions considered in this section, including quantum simulation, drug discovery, quantum machine learning for bioinformatics, sensing, NMR, and imaging.

\subsection{Quantum computing and simulation}

\subsubsection{Quantum simulation of biomolecular systems}

Quantum simulation is motivated by the possibility of representing electronic structure and quantum dynamics more naturally than classical approximations in some regimes, including multireference electronic states that require the simultaneous treatment of several important orbitals (active spaces), as well as dynamical observables for which classical scaling may become prohibitive \cite{kassal2011simulating,Cao2020,Bauer2020,pal2024future,baiardi2023quantum,outeiral2021prospects}. Current work spans algorithm development, proof-of-principle demonstrations, and hybrid quantum--classical workflows for chemistry and dynamics \cite{Bauer2020,navickas2025experimental,fedorov2021towards,andersson2022quantum}. For fault-tolerant or utility-scale applications, a practical route is a hierarchical workflow in which density-functional and other high-level established computational chemistry methods screen large chemical spaces, while quantum phase estimation is reserved for active-space calculations of the most strongly correlated states \cite{Bauer2020,cao2019quantumchem,vonBurg2021Catalysis}. In biological applications, such workflows could improve predictions of reaction energetics, catalytic pathways, and molecular properties relevant to enzyme mechanisms and drug discovery, provided that they demonstrate improved accuracy or cost relative to state-of-the-art classical methods. 

A particularly relevant example for quantum biology is the quantum simulation of radical-pair spin chemistry, where near-term quantum computers have been used to simulate quantum beats and thermal relaxation in radical-pair systems with nontrivial hyperfine interactions \cite{Tolunay2023RadicalPairQC}. This illustrates that quantum simulation is not limited to static molecular energies, but can also target open-system spin dynamics relevant to mechanisms such as radical-pair magnetoreception.

A persistent confound is that classical methods continue to improve rapidly, so claims of advantage depend strongly on baseline choice, target selection, and transparent error accounting; in many realistic settings, approximate classical methods can outperform early quantum prototypes, especially in noisy near-term regimes \cite{cordier2022biology,marx2021biology,cao2019quantumchem}. Decisive progress will therefore require standardized benchmarks with blind targets, explicit cost metrics, and comparison to state-of-the-art classical workflows rather than simplified baselines, together with careful reporting of circuit depth, noise, compilation overhead, data-movement overhead, and end-to-end wall-clock cost in hybrid pipelines \cite{cordier2022biology,marx2021biology,cao2019quantumchem}.

\subsubsection{Quantum computing in molecular design and drug discovery}

Quantum computing is being explored for molecular design, docking, binding estimation, and optimization problems in drug discovery, most often within hybrid workflows that combine classical data-driven models with quantum subroutines for chemistry or search steps \cite{dong2023prediction,malone2022towards,avramouli2024hybrid,domingo2024hybrid,weidner2023leveraging}. At present, however, the key question is not whether such workflows are conceptually plausible, but whether they provide reproducible advantages over strong classical baselines in realistic discovery settings. Major confounds include data quality, target complexity, leakage in retrospective benchmarks, and the possibility that apparent gains disappear once prospective validation, experimental follow-up, and full compute costs are taken into account. Reviews and perspective articles help define the design space, but they do not by themselves constitute evidence of quantum advantage \cite{das2024brief,kumar2024recent,sancho2026npjdrug}. Decisive progress therefore requires prospective predictions with experimental validation, prespecified evaluation metrics, transparent comparison against the best available classical methods, and explicit accounting of compute resources, model-selection latitude, and failure modes \cite{li2021drug}.

\subsubsection{Quantum machine learning and bioinformatics}

Quantum machine learning and quantum bioinformatics aim to address high-dimensional biological data analysis, similarity search, and optimization tasks using quantum or quantum-inspired representations and algorithms \cite{mokhtari2024new,chagneau2024quantum,maheshwari2022quantum,nalkecz2024quantum,schmidt2024gpus,selladurai2024integrating}. A central challenge is that many reported gains are difficult to separate from artifacts of dataset curation, leakage, overfitting, hyperparameter sensitivity, or weak baseline selection, and in many practically accessible regimes the relevant kernels, embeddings, or feature maps may be well approximated classically \cite{cerezo2022qml,bowles2024better,schnabel2025quantumkernel}. Claims of advantage should therefore remain provisional unless they are supported by curated benchmarks with blind splits, replication across platforms, transparent reporting of compute budgets and hyperparameter tuning, and explicit ablations showing which part of the pipeline genuinely depends on a quantum resource rather than on classical preprocessing or model flexibility \cite{cerezo2022qml,bowles2024better,schnabel2025quantumkernel}. The strongest evidence in this area will come not from isolated performance gains on small benchmarks, but from robust demonstrations that quantum methods improve biologically meaningful inference under matched and carefully controlled conditions.

\subsection{Quantum sensing and metrology}

\subsubsection{Diamond quantum sensing in biology}

Nitrogen-vacancy centers in diamond enable detection of magnetic fields and temperature at micro- and nanoscale resolution under ambient conditions \cite{Schirhagl2014,Barry2016,degen2017quantum,taylor2016quantum,kucsko2013nanometre,plakhotnik2014all,gassab2025spin}. The sensing physics is well established, and biological deployments using nanodiamonds, surface functionalization, and scanning or wide-field modalities are expanding rapidly. Practical confounds include calibration drift, surface chemistry and charge-state instability, heating and phototoxicity, motion artifacts in living samples, and the mapping from sensor signals to the biological quantity of interest \cite{aslam2023quantum,davydov2022surface,wu2017nanodiamonds,zhang2021toward,segawa2023nanoscale}. A continuing device-level constraint for wide-field biological imaging is that magnetic-image resolution is often limited by optical diffraction and dense ensemble readout, motivating approaches that seek resolution beyond the optical limit while preserving parallel wide-field acquisition, such as wide-field Fourier magnetic-imaging schemes that encode spatial information in k-space via pulsed gradients and reconstruct super-resolved magnetic maps \cite{guo2024widefieldfourier}. Decisive demonstrations require blind calibration benchmarks, cross-laboratory replication, orthogonal validation against established modalities, and explicit uncertainty propagation from sensor model through calibration to biological inference, together with realistic reporting of throughput and failure rates in heterogeneous biological samples \cite{aslam2023quantum,noureddine2024hybridsensing,davydov2022surface,wu2017nanodiamonds,zhang2021toward}.

\subsubsection{Quantum-enhanced NMR and MRI}

NMR and MRI are limited by thermal polarization. Two quantum-technology routes aim to enhance sensitivity and scale. Diamond quantum sensors enable nano- and microscale NMR detection in regimes inaccessible to conventional coils \cite{allert2022advances,neuling2023prospects,du2024single}. A recent step toward scalable microscopy-style readout is optical wide-field NMR microscopy, in which nitrogen-vacancy ensembles transduce radiofrequency magnetic signals into fluorescence recorded on a high-speed camera, allowing NMR spectra to be acquired in parallel across a field of view rather than by point scanning and enabling multicomponent maps of amplitude, phase, and local field in microstructured samples \cite{briegel2025widefieldnmr}. 

Hyperpolarization methods, by contrast, enhance nuclear-spin polarization by driving spin systems far from thermal equilibrium, thereby increasing magnetic-resonance signal intensity by orders of magnitude and opening routes to improved metabolic imaging, reaction monitoring, and kinetic measurements in biological systems \cite{Eills2023SpinHyperpolarization,jorgensen2022hyperpolarized}. Chemically or photochemically induced dynamic nuclear polarization (CIDNP/photo-CIDNP) provides a compact example at the interface with radical-pair spin chemistry, in which chemical or light-driven radical-pair reactions generate non-Boltzmann nuclear-spin polarization detectable by NMR \cite{Kuhn2013PhotoCIDNPNMR,Javed2025LightActivatedDNP}. For both diamond-sensor NMR and hyperpolarized MRI/NMR, the decisive question is the regime in which sensitivity, spectral resolution, and biocompatible operating conditions can be achieved simultaneously in complex, heterogeneous samples, including constraints imposed by heating, relaxation, surface-induced artifacts, and polarization lifetime. Progress is therefore best judged by benchmark targets with cross-validation to established assays and by explicit accounting of calibration uncertainty and biological variability, consistent with the broader principle that advantage should be defined by inference quality rather than raw sensitivity alone \cite{yukawa2025qls}.

\subsubsection{Quantum-enhanced bioimaging and spectroscopy}

Quantum-light resources, including squeezing and nonclassical photon correlations, enable imaging below shot noise and modalities such as imaging with undetected photons \cite{lemos2014quantum,brida2010experimental}. Biological demonstrations include squeezed illumination, correlation-based concepts for super-resolution and precision enhancement, low-dose imaging regimes, adaptive tracking of enzymatic reactions with quantum light, and theoretical proposals for using quantum states of light to probe retinal-network parameters \cite{taylor2013biological,taylor2014subdiffraction,genovese2016real,moreau2019imaging,Cimini2019EnzymaticQuantumLight,Pedram2022RetinalQuantumLight}. 

Recent work has also begun to evaluate quantum light under explicitly biological constraints, for example by using squeezed light to enhance stimulated Brillouin-scattering contrast in label-free biomechanical imaging while reducing photodamage, with reported gains framed in terms of longer interrogation time and improved sample viability relative to coherent illumination under matched operating conditions \cite{li2024quantumbrillouin}. In parallel, correlated-photon approaches have been used for time- and frequency-resolved optical spectroscopy without ultrashort pulses, exploiting temporal correlations of photon pairs to record fluorescence-lifetime traces in biological samples on sub-second timescales while maintaining wavelength tunability, and to resolve energy-transfer cascades at the single-photon level in photosynthetic membranes \cite{alvarezmendoza2025correlatedphoton}. A central practical question is whether correlation signatures and the resulting inference gains persist in realistic scattering and absorption regimes, since losses and mode mixing can rapidly erase metrological advantage. Recent experiments have begun to probe this robustness in scattering media and biological material, including dose-constrained regimes \cite{shi2016photon,galvez2022decoherence,zhang2024quantum}. Major confounds include detector losses, classical technical noise, stability requirements, and restriction to niche operating regimes. Decisive comparisons should therefore hold dose and acquisition time fixed, benchmark against strong classical modalities, report reproducibility across laboratories and sample variability, and make clear whether the quantum resource improves an end biological inference task rather than only an intermediate optical metric \cite{moreau2019imaging,zheltikov2020photon,bowen2023quantum}.

\subsubsection{Quantum pupillometry and retinal detection maps}

A related visual-measurement direction is quantum pupillometry and retinal biometrics, where individual-specific detection-efficiency maps are treated as possible sources of biometric or visual-function contrast \cite{tinsley2016direct,holmes2017temporal,hecht1942energy,rieke1998single,loulakis2017quantum,margaritakis2020stimulus}. This is best understood as a quantum-for-biology application rather than as a mechanistic claim that retinal physiology requires nonclassical dynamics. The relevant evidential standard is therefore translational: whether near-quantum-limited visual measurements improve biological or biometric inference under realistic protocols. Dominant confounds include physiological variability, attention and adaptation effects, protocol standardization, device dependence, and the security and privacy constraints associated with deploying eye- and pupil-based biometrics at scale \cite{morales2019irispad,liebling2014privacy,liu2019eyeprivacy}. Decisive tests require large-cohort blinded studies with prespecified protocols, robust estimation of error rates in realistic conditions, and explicit privacy and presentation-attack models tied to device constraints and population variability.

\subsection{Quantum-inspired decision theory}

Quantum-probability models apply the formal structure of quantum theory to cognition and decision making without implying that the brain physically instantiates quantum states \cite{khrennikov2009quantum,khrennikov2015quantum,busemeyer2012quantum,pothos2022quantum}. Their main value lies in providing compact mathematical descriptions of contextuality, order effects, and interference-like patterns in behavior that are often difficult to capture within standard classical probabilistic frameworks \cite{pothos2022quantum,wang2014context}. The central confound is therefore not experimental noise alone, but category confusion between useful formal analogy and physical mechanism, together with model flexibility that can fit existing data without yielding distinctive predictions. Stronger evidence in this area would come from preregistered out-of-sample prediction and rigorous model comparison that penalizes flexibility, evaluates generalization, and remains explicit that success at the level of formal description does not by itself constitute evidence for physical quantum biology.

\subsection{Section synthesis}

Within quantum for biology, the strongest near-term case lies in sensing, metrology, and selected imaging modalities, where the quantum hardware is already real and the open question is whether the full measurement-to-inference chain outperforms strong classical alternatives in practice. Quantum simulation, drug discovery, and quantum machine learning remain strategically important, but the appropriate standard there is benchmark discipline rather than premature claims of biological advantage. Quantum-inspired decision-theoretic and cognitive models are useful as mathematical frameworks when they predict better than classical alternatives under strict model comparison, but they should not be conflated with evidence that brains instantiate physical quantum states \cite{degen2017quantum,moreau2019imaging,Cao2020,sancho2026npjdrug,pothos2022quantum,busemeyer2012quantum}. A compact topic-by-topic summary of this section is provided in Appendix Table~\ref{tab:map_qfb}.


\section{Biology for Quantum}

\begin{figure*}[t]
    \centering
    \includegraphics[width=\textwidth]{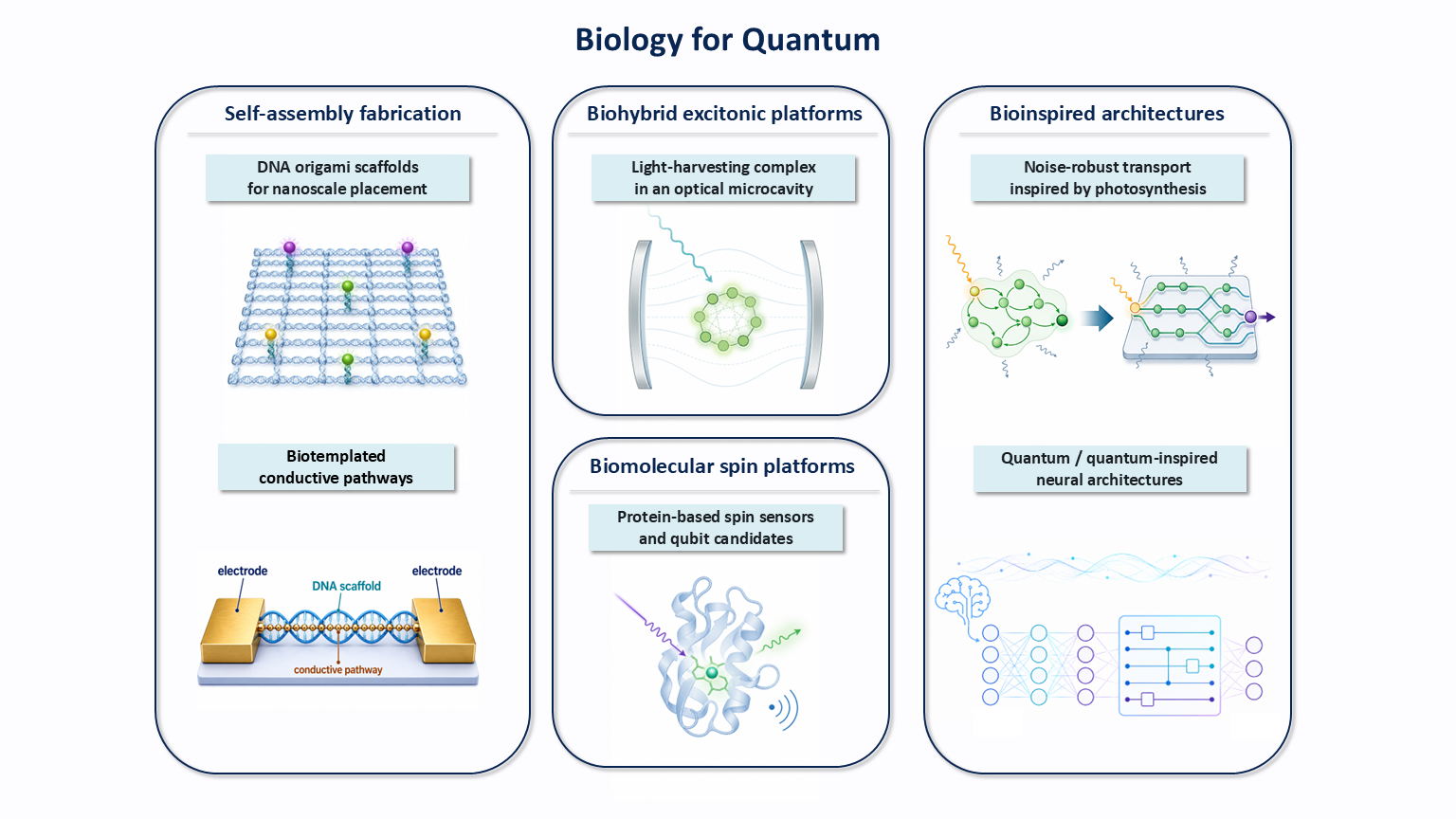}
    \caption{Overview of representative ways in which biological structures, materials, and design principles may inform or support quantum technologies. The panels schematically illustrate self-assembly and fabrication strategies, including DNA origami scaffolds for nanoscale placement and biotemplated conductive pathways; biohybrid excitonic platforms, including light-harvesting complexes in optical microcavities; biomolecular spin platforms, including protein-based spin sensors and qubit candidates; and bioinspired architectures, including noise-robust transport inspired by photosynthesis and quantum or quantum-inspired neural-network architectures. These examples illustrate possible routes by which biological organization or function may be translated into quantum-device concepts, with evidential strength depending on device-level benchmarks, reproducibility, and comparison with non-biological alternatives.}    
    \label{fig:biology_for_quantum_overview}
\end{figure*}

Biology for quantum concerns biological principles and biomolecular architectures that can inspire or enable quantum technologies. Biological inspiration is most compelling when it yields measurable improvements in device or algorithmic performance under realistic conditions of noise, disorder, and fabrication variability. The central question is therefore not whether biology is complex or self-organized, but whether it provides engineering capabilities that are difficult to achieve using conventional nonbiological design strategies. In this section, the relevant evidence ladder is therefore primarily technological rather than physiological: the strongest claims are those in which a biological architecture measurably improves fabrication, integration, robustness, sensing, or device performance relative to a nonbiological alternative. Figure~\ref{fig:biology_for_quantum_overview} summarizes the main directions discussed in this section, with emphasis on areas in which biology may contribute distinctive capabilities such as programmable self-assembly, genetic encodability, soft-matter integration, and robustness to structural variability.

\subsection{Self-assembly as fabrication for quantum technologies}

\subsubsection{DNA origami scaffolds for quantum nanophotonics}

DNA origami enables programmable self-assembly and nanometer-scale placement of functional components, including quantum emitters and plasmonic structures, and has emerged as a useful platform for organizing nanoscale photonic architectures \cite{kuzyk2018dnaorigami,hong2017dnaorigami,zhan2023dnaorigami}. The main biological contribution here is addressable and modular self-assembly, which can be used to position interacting components with high spatial control \cite{kuzyk2018dnaorigami,hong2017dnaorigami,zhan2023dnaorigami}. The main uncertainties concern stability in operational environments, integration with photonic platforms, emitter photophysics near metals or surfaces, and the extent to which assembly variability dominates device performance. Decisive benchmarks therefore require reproducibility across large numbers of assembled structures, quantitative characterization of optical-coupling distributions and failure modes, and direct comparison with top-down alternatives while holding the photonic context fixed so that the contribution of biological self-assembly is isolated.

\subsubsection{Biotemplated bottom-up fabrication}

Biological templates such as viruses, protein assemblies, and DNA scaffolds offer bottom-up routes to organizing inorganic nanowires, conductive pathways, and hybrid nanostructures with nanoscale addressability \cite{mao2004virusnanowires,braun1998dnaSilverWire}. In the context of quantum technology, the key question is not simply whether such structures can be assembled, but whether the biological route yields lower disorder, better scalability, or functionalities that are difficult to reproduce using conventional fabrication. These approaches are conceptually attractive because they can position and connect nanoscale elements under mild conditions, but their relevance to quantum devices depends on whether they can ultimately satisfy stricter requirements on purity, surface loss, interface quality, and reproducibility. Decisive progress therefore requires direct benchmarking of optical or electronic loss, noise, and variability relative to conventional fabrication strategies, rather than proof-of-principle demonstrations of connectivity or templating alone.

Ferritin is also relevant as a protein-cage platform for quantum-adjacent device concepts, because its iron-oxide core, self-assembled shell, and reported electron-transport behavior make it a candidate biological template for nanoscale electronic, magnetic, and sensing architectures \cite{DiezPerez2023FerritinTunneling,Kumar2016FerritinTunneling,Bera2019FerritinMultilayers}.

\subsection{Biohybrid excitonic and polaritonic platforms}

Strong coupling between biological light-harvesting structures and confined optical modes has been demonstrated, including coupling between chlorosomes of photosynthetic bacteria and confined optical cavity modes \cite{coles2014chlorosomes}. Such work motivates biohybrid polaritonic platforms in which biological chromophore networks can contribute large oscillator strengths, rich vibronic structure, and self-organized architectures within the broader landscape of room-temperature molecular strong coupling and polaritonic photonics \cite{coles2014chlorosomes,xiang2024molecularpolaritons,hertzog2019strong}. For biology for quantum, the key question is whether these components enable device regimes that are difficult to access with synthetic materials alone, such as robust room-temperature collective coupling, tunable polaritonic dispersion, or disorder-tolerant optical functionality. Confounds include sample heterogeneity, photobleaching, and the fact that spectroscopic signatures of strong coupling do not by themselves establish useful quantum coherence, controllability, or device-level advantage \cite{xiang2024molecularpolaritons,perezsanchez2024collective}. Decisive benchmarks therefore focus on stability, reproducible coupling strengths, tunability, and integration into photonic circuits where performance can be assessed end to end.

\subsection{Biomolecular spin platforms and genetically encoded qubits}

A distinct emerging direction is the embedding of quantum-relevant spin degrees of freedom in biomolecular scaffolds. A striking example is the demonstration of a fluorescent-protein spin qubit platform, which uses a biomolecule to host and protect a controllable quantum state under aqueous and ambient-compatible conditions \cite{feder2025fluorescentqubit}. A closely related advance showed that engineered magneto-sensitive fluorescent proteins can exhibit optically detected magnetic resonance in living cells at room temperature, and that directed evolution can tune magnetic- and radiofrequency-response properties, supporting a path toward genetically encodable and potentially in vivo-compatible quantum sensing architectures \cite{abrahams2026quantumspinresonance}. These results do not imply that biology naturally performs quantum computation, but they do support a narrower biology-for-quantum claim: biomolecular structure, solubility, and genetic encodability may provide routes to scalable fabrication, targeting, and environmental integration of spin centers. Key engineering questions now include whether coherence times and control fidelity remain competitive as biological complexity increases, whether photochemical stability can be maintained under repeated control cycles, and how inhomogeneity across expression batches affects device metrics. Decisive progress is therefore defined by standard qubit and sensing benchmarks, including coherence times under relevant conditions, gate or control fidelities, integration into sensing or networking protocols, and reproducibility across expression contexts.

\subsection{Bioinspired methods and architectures}

\subsubsection{Noise-robust transport and interference inspired by photosynthesis}

Photosynthetic light harvesting motivates engineered designs in which environmental noise and interference may be used constructively in transport networks \cite{plenio2008dephasing,Creatore2013,Romero2014,sarovar2010quantum}. The biological contribution lies in showing that functional efficiency can coexist with disorder and strong environmental coupling when architecture and energetics are appropriately structured, thereby suggesting design rules for robustness rather than only idealized optimality. For quantum technologies, the relevant benchmarks are therefore robustness and scalability: improvements that persist across disorder ensembles and fabrication variability, together with explicit comparison to strong classical designs under matched noise and disorder assumptions.

\subsubsection{Quantum biomimetics and artificial-life-inspired algorithms}

Quantum biomimetics provides another bioinspired route for quantum technologies, but its status should be treated carefully. In this context, the goal is not to claim that living systems are quantum artificial-life machines, but to use selected biological behaviors, such as self-replication, mutation, interaction, inheritance, adaptation, and death, as design motifs for quantum-information protocols and controllable toy models \cite{AlvarezRodriguez2014BiomimeticCloning,AlvarezRodriguez2016ArtificialLifeQT,AlvarezRodriguez2018QuantumArtificialLife,Lamata2020QMLBiomimetics}. Proof-of-principle work has shown that such behaviors can be encoded in quantum platforms, including an implementation of a quantum artificial-life algorithm on an IBM quantum computer, while related models have explored photon-driven self-replication and links to dissipative adaptation \cite{AlvarezRodriguez2018QuantumArtificialLife,Valente2021SelfReplication}. The evidential status is therefore exploratory and design-motivating: these studies extend the repertoire of bioinspired quantum algorithms and simulations, but they do not by themselves establish a biological quantum mechanism or a practical quantum advantage. Decisive progress would require benchmarked tasks in which artificial-life-inspired quantum protocols outperform strong classical artificial-life, evolutionary, or standard quantum-algorithm baselines under transparent resource accounting.

\subsubsection{Brain-inspired quantum and quantum-inspired neural computation}

Brain-inspired computing draws on principles of biological neural systems in both algorithms and hardware, including event-driven dynamics, temporal coding, and energy-efficient information processing \cite{Kudithipudi2025}. Models span a range of abstraction, from biophysically detailed descriptions such as Hodgkin--Huxley dynamics to simplified spiking models such as leaky integrate-and-fire neurons \cite{Chen2026,Liu2026neural}. These form the basis of spiking neural networks, which encode information in the timing of discrete spikes and are often motivated by efficiency and temporal processing rather than by biological realism alone \cite{Kudithipudi2025,Chen2026,Liu2026neural}.

A related emerging direction is quantum neuromorphic computing, which seeks to combine brain-inspired computational architectures with quantum resources such as superposition, interference, and entanglement \cite{markovic2020quantum,pfeiffer2016quantum,salmilehto2017quantum,prati2017quantum,sanz2018quantum,guo2021quantum,kristensen2021spiking,brand2024quantum,stremoukhov2026quantum}. At present, however, this remains largely theoretical or proof-of-principle. The key question is therefore not whether such architectures are conceptually interesting, but whether they provide advantages over strong classical neuromorphic systems or more standard quantum-computing approaches on well-defined tasks. Decisive evaluation requires matched benchmarks for scalability, robustness to noise, resource demands, and performance on representative problem classes, together with comparison to classical deep learning, classical neuromorphic hardware, and established quantum algorithms.

Quantum-inspired neural computation uses Hilbert-space representations, quantum probability, contextuality, and quantum-information metaphors to design models and algorithms for complex data, including cognition and neuroinformatics \cite{PothosBusemeyer2013,Wang2013QuantumCognition,Orka2025QuantumNeuroinformatics}. These approaches should be understood as formal or computational models rather than evidence that the brain operates through microscopic quantum coherence. A complementary theoretical direction argues that large classical synchronized networks can generate emergent quantum-like state spaces and interference-like probability laws, and discusses possible relevance to brain function as an open question rather than a demonstrated mechanism \cite{Scholes2024QLNetworks}. More explicitly brain-directed quantum-like modeling has also proposed bridges from oscillatory neuronal network activity to quantum-like cognitive descriptions, including frameworks for modeling ``mental entanglement'' at the level of network representations, while emphasizing that empirical validation would require carefully designed EEG- or MEG-based tests \cite{KhrennikovYamada2025MentalEntanglement}. The main confound is that classical baselines are already strong and continually improving, so claims of advantage require careful benchmarking with matched compute budgets, control of leakage and dataset bias, and rigorous evaluation of generalization and robustness across cohorts and acquisition settings.

\subsection{Section synthesis}

Biology for quantum is strongest when the biological contribution is not merely illustrative but difficult to replace by conventional design strategies, and when that contribution is measured through explicit device, fabrication, sensing, or algorithmic benchmarks. DNA origami and related self-assembly platforms already provide concrete enabling capabilities for nanoscale positioning and fabrication. Biohybrid polaritonic systems and biomolecular spin platforms are scientifically credible emerging directions, but they still require stronger benchmarks for stability, reproducibility, and device-level performance. By contrast, quantum biomimetic and artificial-life-inspired protocols, together with broader brain-inspired or quantum-like architectures, should presently be treated as hypothesis-generating and design-motivating unless they can be tied to explicit hardware, sensing, or algorithmic advantages under matched comparisons \cite{kuzyk2018dnaorigami,feder2025fluorescentqubit,abrahams2026quantumspinresonance,AlvarezRodriguez2016ArtificialLifeQT,AlvarezRodriguez2018QuantumArtificialLife,Lamata2020QMLBiomimetics,Scholes2024QLNetworks}. A compact topic-by-topic summary of this section is provided in Appendix Table~\ref{tab:map_bfq}.

\section{Development of the Field and Future Perspectives}
Across all three directions, progress is most reliable when claims are mechanistically specific and tested in genuinely discriminating ways. In quantum in biology, the most mature cases remain those tied to well-constrained degrees of freedom and measurable kinetics, such as tunneling contributions in enzymatic hydrogen transfer and the spin-chemistry framework underlying radical-pair models, whereas areas such as microtubule photophysics remain scientifically interesting but unresolved and should not presently be cited as settled physiological mechanisms \cite{kalra2023electronic,Marais2018}. In quantum for biology, advantage must be defined operationally through transparent benchmarks and uncertainty budgets, because classical baselines continue to improve rapidly and biological inference remains highly sensitive to calibration, noise, and confounds \cite{cordier2022biology,marx2021biology,degen2017quantum,moreau2019imaging}. In biology for quantum, biological inspiration is most compelling when it yields approaches that remain effective across disorder and variability, or when it enables fabrication strategies that are difficult to reproduce by conventional means while still meeting quantum-device performance criteria \cite{kuzyk2018dnaorigami,Creatore2013,mao2004virusnanowires}.

A practical road ahead follows three closely linked principles. First, experimental validation should prioritize perturbation and scaling tests that isolate the proposed quantum ingredient while tightly controlling classical alternatives, and replication across laboratories should be treated as a primary evidential criterion rather than as a secondary follow-up step \cite{ball2011physics,Marais2018}. Recent microtubule experiments based on magnetic-isotope perturbations illustrate the kind of more discriminating strategy that can move a topic away from broad speculation and toward experimentally constrained mechanism \cite{ZadehHaghighi2026}. Second, predictive models and technology claims should be evaluated using transparent benchmarks and explicit uncertainty propagation, so that improvements in sensors, simulators, and algorithms translate into improved biological inference rather than merely into more precise but insufficiently validated intermediate signals \cite{degen2017quantum,moreau2019imaging}. Third, multiscale theory should connect microscopic dynamics to biological outputs and device performance by explicitly linking modeling regimes, from electronic-structure calculations and QM/MM (quantum mechanics/molecular mechanics) embedding, through atomistic and coarse-grained molecular dynamics, to open quantum-system descriptions of coherent dynamics, relaxation, and decoherence where these are mechanistically relevant \cite{Marais2018,Gerhards2026Multiscale,BreuerPetruccione2002Open}. Such models should state assumptions, propagate uncertainties, and generate falsifiable predictions, allowing the evidence map presented in this review to evolve as decisive tests accumulate rather than as narratives become more elaborate.

A further long-term implication concerns altered magnetic environments. Because hyperfine interactions are set by local molecular and protein environments, and because Earth-evolved biology has developed under the geomagnetic field, prolonged hypomagnetic or otherwise altered magnetic exposure, for example during space travel or planetary settlement, could in principle perturb some spin-dependent reaction yields or downstream redox regulation. At present, this should be treated as a speculative but testable extension of radical-pair biology rather than an established physiological risk \cite{ZadehHaghighi2023Hypomagnetic,Adams2026QuantumEvolution}.

A useful discipline across this interface is to require every strong claim to answer five questions explicitly: What specific quantum resource is being invoked? What is the minimal mechanistic or algorithmic model that generates distinctive predictions? What is the strongest classical alternative under the same conditions? What perturbation, benchmark, or matched comparison would discriminate between them? And what biological or engineering output would count as success? Claims that cannot yet be mapped onto this framework are often best presented as hypotheses or design directions rather than as established results. This caution is particularly important in adjacent areas such as photobiomodulation and light-responsive mitochondrial signaling, where reported biological effects are clearly of interest, even though the full mechanistic chain remains incompletely resolved and is unlikely to be captured by a single explanatory model \cite{Sanderson2018NIR,Liebert2023PBM,Arany2014TGFb}. Similarly, work on ultra-weak photon emission highlights the importance of distinguishing experimentally established photophysical signals, whose intensity, spectrum, and timing may provide multidimensional redox-metabolic readouts, from stronger claims about coherence or nonclassicality, for which reliable evidence remains limited \cite{Cifra2015BiophotonsCoherence}.

The same discipline also sharpens the selective agenda of this review. For quantum in biology, the most important next step is not broader speculation, but more discriminating perturbation experiments in systems where mechanistic alternatives can genuinely be separated. For quantum for biology, the most important next step is benchmark discipline: blind targets, matched dose or cost conditions, uncertainty budgets, and biologically meaningful endpoints rather than surrogate technical metrics. For biology for quantum, the most important next step is to show that biological structure provides engineering capabilities that survive manufacturing variability, device integration, and long-term reproducibility tests. Under these standards, some subfields already appear durable, some appear promising, and some should still be described as exploratory.

Ultimately, the contribution of this review is best understood as a decision framework for future work rather than as an attempt to maximize topic count. The field is likely to advance fastest where mechanistic precision, benchmark discipline, and engineering realism remain tightly coupled.

\section{Acknowledgments}
During the preparation of this manuscript, GPT-assisted tools were used in a limited editorial role to support clarity, organization, and language refinement. Selected schematic visual elements were also generated or refined using GPT-assisted image tools, with final figures assembled and edited by the authors. All scientific content, literature selection, interpretation, figure selection, and final wording were independently reviewed and approved by the authors. The authors take full responsibility for the accuracy, integrity, and originality of the manuscript.

This research was enabled in part by support provided by the Digital Research Alliance of Canada to T.J.A.C. This research was undertaken in part thanks to funding to T.J.A.C. from the Canada Research Chairs Program (CRC-2022-00204) and the University of Waterloo. L.G. acknowledges support from the University of Waterloo Provost’s Program for Interdisciplinary Postdoctoral Scholars. B.A. was supported by the National Institute for Theoretical and Computational Sciences.

The authors thank Chris Rourk for his thoughtful feedback and for highlighting ferritin-related literature that helped improve the discussion of biomolecular electron tunneling, nanomagnetism, and possible bioelectric or biomagnetic mechanisms.

\appendix
\section{Summary tables across the map}

This Appendix makes the review's evidential framework explicit. Table~\ref{tab:evidence_tiers} provides a compact cross-map maturity rubric, while Tables~\ref{tab:map_qib}--\ref{tab:map_bfq} summarize the main claims, confounds, and decisive next tests for each direction.

\begin{table*}[t]
\caption{Cross-map evidence tiers used throughout the review. This compact rubric summarizes the review's comparative judgments before the more detailed navigation tables.}
\label{tab:evidence_tiers}
\begin{adjustbox}{width=\textwidth}
\begin{tabular}{llll}
\toprule
\mapcell{0.15\textwidth}{Direction} & \mapcell{0.24\textwidth}{Topic cluster} & \mapcell{0.31\textwidth}{Present status} & \mapcell{0.26\textwidth}{Most efficient upgrade path} \\
\midrule
\mapcell{0.15\textwidth}{Quantum in Biology} & \mapcell{0.24\textwidth}{Molecular quantum chemistry and transformation} & \mapcell{0.31\textwidth}{Most secure where microscopic quantum-chemical effects are tightly constrained by kinetics and mechanism, especially enzymatic hydrogen tunneling and related PCET chemistry} & \mapcell{0.26\textwidth}{Link microscopic transfer mechanisms more tightly to full catalytic cycles, polymerase-resolved outcomes, and perturbation-resolved biological phenotypes} \\\addlinespace

\mapcell{0.15\textwidth}{Quantum in Biology} & \mapcell{0.24\textwidth}{Sensory transduction} & \mapcell{0.31\textwidth}{Strongest for photon detection and radical-pair spin chemistry; olfaction remains more theoretical and receptor-level evidence is limited} & \mapcell{0.26\textwidth}{Use receptor-, protein-, and organism-level perturbations tied directly to sensory outputs, with controls for classical binding, adaptation, and transduction effects} \\\addlinespace

\mapcell{0.15\textwidth}{Quantum in Biology} & \mapcell{0.24\textwidth}{Excited-state, redox, and spin transfer} & \mapcell{0.31\textwidth}{Strong physicochemical or ultrafast evidence in several systems, but functional necessity under biological conditions remains conditional and perturbation-dependent} & \mapcell{0.26\textwidth}{Use dephasing-, detuning-, redox-, spin-, or photochemical perturbations tied directly to yield, signaling, repair, or bioenergetic output} \\\addlinespace

\mapcell{0.15\textwidth}{Quantum in Biology} & \mapcell{0.24\textwidth}{Macromolecular transport and collective effects} & \mapcell{0.31\textwidth}{Experimentally and theoretically active, especially for microtubule photophysics and spin-sensitive tubulin effects, but not yet established as physiological quantum mechanisms} & \mapcell{0.26\textwidth}{Use independent replication, standardized preparations, explicit sample-quality controls, and observables that distinguish coherent delocalization, resonant transport, or spin-sensitive effects from classical hopping or preparation artifacts} \\\addlinespace

\mapcell{0.15\textwidth}{Quantum in Biology} & \mapcell{0.24\textwidth}{Information Processing} & \mapcell{0.31\textwidth}{More formal or speculative at present, with the main gap being direct evidence for physiological nonclassical states or predictive advantage beyond classical alternatives} & \mapcell{0.26\textwidth}{Use direct cluster or state detection where possible, physiological tests of spin relaxation and coherence, and strict model-comparison benchmarks for information-processing formalisms} \\\addlinespace

\mapcell{0.15\textwidth}{Quantum for Biology} & \mapcell{0.24\textwidth}{Diamond quantum sensing} & \mapcell{0.31\textwidth}{Technologically mature device physics; the biological inference chain remains the limiting step} & \mapcell{0.26\textwidth}{Require cross-laboratory calibration, uncertainty propagation, and end-task biological validation} \\\addlinespace

\mapcell{0.15\textwidth}{Quantum for Biology} & \mapcell{0.24\textwidth}{Quantum-enhanced NMR/MRI and imaging} & \mapcell{0.31\textwidth}{Rapidly emerging, with several credible methodological advances and a clear benchmark structure} & \mapcell{0.26\textwidth}{Use dose-, time-, and sample-matched comparisons against the strongest classical modalities} \\\addlinespace

\mapcell{0.15\textwidth}{Quantum for Biology} & \mapcell{0.24\textwidth}{Quantum simulation, drug discovery, and quantum machine learning} & \mapcell{0.31\textwidth}{Promising but still benchmark-limited; strong claims should remain prospective} & \mapcell{0.26\textwidth}{Use blind targets, prospective validation, transparent compute accounting, and stronger classical baselines} \\\addlinespace

\mapcell{0.15\textwidth}{Biology for Quantum} & \mapcell{0.24\textwidth}{DNA origami and self-assembly platforms} & \mapcell{0.31\textwidth}{Established enabling route for positioning and fabrication} & \mapcell{0.26\textwidth}{Demonstrate large-scale reproducibility within integrated quantum-device workflows} \\\addlinespace

\mapcell{0.15\textwidth}{Biology for Quantum} & \mapcell{0.24\textwidth}{Biohybrid polaritons and biomolecular spin platforms} & \mapcell{0.31\textwidth}{Scientifically credible emerging device directions} & \mapcell{0.26\textwidth}{Require stability, fidelity, coherence, and reproducibility benchmarks under operating conditions} \\\addlinespace

\mapcell{0.15\textwidth}{Biology for Quantum} & 
\mapcell{0.24\textwidth}{Bioinspired transport, artificial-life, and neural architectures} & \mapcell{0.31\textwidth}{Conceptually suggestive, but often still closer to design inspiration than to measured quantum advantage} & \mapcell{0.26\textwidth}{Show device- or algorithm-level gains that survive matched classical comparisons} \\
\bottomrule
\end{tabular}
\end{adjustbox}
\end{table*}

\begin{table*}[t]
\caption{Navigation table for Quantum in Biology. Rows summarize the central claim, the strongest evidence base, key confounds, and the most decisive tests that would change confidence.}
\label{tab:map_qib}
\begin{adjustbox}{width=\textwidth}
\begin{tabular}{lllll}
\toprule
\mapcell{0.15\textwidth}{Topic} & \mapcell{0.25\textwidth}{Central claim} & \mapcell{0.24\textwidth}{Strongest evidence base} & \mapcell{0.24\textwidth}{Main confounds} & \mapcell{0.30\textwidth}{Decisive tests}\\
\midrule
\mapcell{0.15\textwidth}{DNA replication} & \mapcell{0.25\textwidth}{Tautomerization and mispairing may be influenced by proton tunneling or proton delocalization \cite{loewdin1963proton,pusuluk2018quantum}} & \mapcell{0.24\textwidth}{Quantum-chemistry plausibility and mechanistic modeling \cite{pusuluk2018quantum}} & \mapcell{0.24\textwidth}{Polymerase conformational selection, proofreading, and classical fluctuations} & \mapcell{0.30\textwidth}{Isotope and temperature scaling in polymerase-resolved assays with tautomer-lifetime and misincorporation constraints}\\\addlinespace

\mapcell{0.15\textwidth}{Enzyme catalysis} & \mapcell{0.25\textwidth}{Hydrogen transfer includes tunneling contributions \cite{klinman2013hydrogen}} & \mapcell{0.24\textwidth}{Kinetic isotope effects, temperature trends, pressure or viscosity perturbations, and constrained theory} & \mapcell{0.24\textwidth}{Coupled catalytic steps, gating motions, and shifting rate limitation} & \mapcell{0.30\textwidth}{Targeted substitution or directed-evolution series with constrained modeling and orthogonal structural constraints \cite{korchagina2025directed}}\\\addlinespace

\mapcell{0.15\textwidth}{PCET motifs} & \mapcell{0.25\textwidth}{Coupled electron--proton transfer governs key redox and proton-transfer steps \cite{Mayer2004ARPC,HammesSchiffer2010ChemRev}} & \mapcell{0.24\textwidth}{Physical-chemistry theory and targeted model systems} & \mapcell{0.24\textwidth}{Attribution within full catalytic cycles and competition from semiclassical rate descriptions} & \mapcell{0.30\textwidth}{Isotope, pH, and electrostatic scaling in isolated biological steps while holding the redox landscape as constant as possible}\\\addlinespace

\mapcell{0.15\textwidth}{Vision} & \mapcell{0.25\textwidth}{Near-quantum-limited photon detection and ultrafast photochemical dynamics in rhodopsin \cite{hecht1942energy,rieke1998single}} & \mapcell{0.24\textwidth}{Single-photon physiology, ultrafast photochemistry, and quantum-resource-theoretic analyses} & \mapcell{0.24\textwidth}{Coherence signatures can be mimicked by vibrational wavepackets, ensemble effects, and excitation-condition dependence} & \mapcell{0.30\textwidth}{Controlled dephasing or vibronic perturbations linked to quantum yield, isomerization timing, or downstream amplification statistics}\\\addlinespace

\mapcell{0.15\textwidth}{Olfaction and vibrationally assisted receptor activation} & \mapcell{0.25\textwidth}{Vibrationally assisted electron transfer may contribute to receptor activation \cite{bittner2012quantum,solov2012vibrationally,tirandaz2017validity}} & \mapcell{0.28\textwidth}{Theoretical and open-system plausibility under receptor-like dissipative conditions \cite{tirandaz2017validity,checinska2015dissipation}} & \mapcell{0.28\textwidth}{Binding and dynamical explanations; limited receptor-level evidence \cite{block2015implausibility,hoehn2018status}} & \mapcell{0.30\textwidth}{Receptor-resolved isotopologue tests controlling affinity and binding pose, together with detection of charge-transfer signatures}\\\addlinespace

\mapcell{0.15\textwidth}{Magnetoreception} & \mapcell{0.25\textwidth}{Radical-pair spin dynamics can yield magnetic sensitivity \cite{Ritz2000,cai2013chemical}} & \mapcell{0.28\textwidth}{Open-system spin models, chemically constrained estimates, and proxy spin-chemistry experiments \cite{adams2018open,pearson2016experimental}} & \mapcell{0.28\textwidth}{Species dependence, competing mechanisms, radical identity, and transduction uncertainty \cite{xie2022searching}} & \mapcell{0.30\textwidth}{Genetics, spectroscopy, identified radicals, and controlled electromagnetic perturbations tied to predicted behavioral or physiological effects}\\\addlinespace

\mapcell{0.15\textwidth}{Photosynthesis} & \mapcell{0.25\textwidth}{Coherence-like or vibronically mixed excitonic dynamics may influence energy transport \cite{Engel2007,Panitchayangkoon2010,Collini2010}} & \mapcell{0.28\textwidth}{Reproducible ultrafast signatures and open-system theory \cite{ishizaki2009theoretical,chen2015using}} & \mapcell{0.28\textwidth}{Vibrational mimicry, ensemble effects, static disorder, and difficulty proving in vivo functional advantage \cite{runeson2022explaining}} & \mapcell{0.30\textwidth}{Selective detuning or dephasing perturbations linked to functional yield under near-native conditions}\\\addlinespace

\mapcell{0.15\textwidth}{DNA charge transport} & \mapcell{0.25\textwidth}{DNA-mediated redox transport may contribute to lesion sensing or redox signaling \cite{Merino2008DNACT,Sontz2012ACCR}} & \mapcell{0.28\textwidth}{Physicochemical evidence and repair-protein redox studies \cite{Merino2008DNACT,Sontz2012ACCR}} & \mapcell{0.28\textwidth}{Competing protein recruitment, diffusive search, chromatin context, and lesion-recognition steps may dominate repair kinetics} & \mapcell{0.30\textwidth}{Nucleosome- or chromatin-like substrate tests and cell-based assays with redox mutants plus quantitative model comparison}\\\addlinespace

\mapcell{0.15\textwidth}{DNA repair photochemistry} & \mapcell{0.25\textwidth}{Photolyase uses excited-state photochemistry and ultrafast electron transfer to repair lesions \cite{Sancar2003Photolyase,Thiagarajan2011PNAS}} & \mapcell{0.28\textwidth}{Mechanistic photochemistry, identified cofactors, and repair kinetics} & \mapcell{0.28\textwidth}{Stronger coherence claims remain difficult to isolate in fluctuating protein environments and depend on excitation conditions} & \mapcell{0.30\textwidth}{Perturbations that selectively alter dephasing or related excited-state dynamics and test corresponding changes in repair yield or pathway branching}\\\addlinespace

\mapcell{0.15\textwidth}{Cellular respiration} & \mapcell{0.25\textwidth}{Tunneling corrections may influence electron or proton transfer in respiratory complexes \cite{moser2006electron,hayashi2011quantum}} & \mapcell{0.28\textwidth}{Mechanistic modeling, transfer-pathway calculations, and biochemical constraints} & \mapcell{0.28\textwidth}{Classical transfer with gating, reorganization, membrane-potential effects, and redox-landscape changes} & \mapcell{0.30\textwidth}{Perturbations isolating tunneling pathways with orthogonal biochemical validation and controlled redox conditions}\\\addlinespace

\mapcell{0.15\textwidth}{Ultra-weak photon emission} & 
\mapcell{0.25\textwidth}{Biological systems emit ultra-weak photons linked to oxidative metabolism, ROS chemistry, electronically excited species, and mitochondrial redox activity \cite{PospisilPrasadRac2019ExcitedSpecies,Mould2024UPEReview,VanWijk2020MitochondrialUPE}} & 
\mapcell{0.24\textwidth}{Recent studies support UPE as a possible marker of biological state and selected radiation-bystander or mitochondrial responses \cite{Murugan2020CancerUPE,Belksma2026IschemiaReperfusionUPE,Le2017BiophotonExosomes,Tong2024BiophotonSignaling,Mould2023NonChemicalMitochondria}} & 
\mapcell{0.24\textwidth}{Photon emission alone does not demonstrate coherence, nonclassical light, functional optical communication, or source-specific metabolic inference \cite{Cifra2015BiophotonsCoherence}} & 
\mapcell{0.30\textwidth}{Calibrated spectral, temporal, and photon-statistical measurements with independent ROS, metabolic, electrophysiological, optical-transport, and tissue-calibration controls}\\\addlinespace

\mapcell{0.15\textwidth}{CISS} & \mapcell{0.25\textwidth}{Chiral biomolecular scaffolds could influence spin selectivity in electron transfer and downstream spin-sensitive chemistry \cite{bloom2024chemrev,aiello2022chirality}} & \mapcell{0.28\textwidth}{Strong interface and junction evidence for CISS itself \cite{bloom2024chemrev,aiello2022chirality}} & \mapcell{0.28\textwidth}{Interfacial artifacts, contact geometry, and classical parameter changes} & \mapcell{0.30\textwidth}{Solution-phase or near-physiological tests with chirality-matched controls and independent spin-sensitive readouts}\\\addlinespace

\mapcell{0.15\textwidth}{Ferritin protein nanoparticles} & \mapcell{0.25\textwidth}{Ferritin cores and protein shells may support electron tunneling, Coulomb-blockade-like behavior, and magnetic effects relevant to bioelectric or biomagnetic hypotheses \cite{DiezPerez2023FerritinTunneling,Rourk2025FerritinBioelectricity}} & \mapcell{0.28\textwidth}{Ferritin junction and multilayer transport experiments, together with review-level synthesis of ferritin electrical and magnetic properties \cite{Kumar2016FerritinTunneling,Bera2019FerritinMultilayers,Rourk2021FerritinDMFS}} & \mapcell{0.28\textwidth}{Device geometry, iron loading, hydration, aggregation, redox state, cellular localization, and uncertain linkage to biological function} & \mapcell{0.30\textwidth}{In situ or near-native perturbation tests controlling iron loading, redox state, magnetic field, chirality or spin-sensitive readouts, and defined biological endpoints}\\\addlinespace

\mapcell{0.15\textwidth}{Ion channels} & \mapcell{0.25\textwidth}{Coherent-transport hypotheses in selectivity filters could contribute to selectivity or conduction under some conditions \cite{VaziriPlenio2010NJP,Ganim2011NJP}} & \mapcell{0.28\textwidth}{Theoretical proposals; established quantum chemistry of ion energetics and proton pathways \cite{Roux2011IonSelectivity,DeCoursey2003ProtonChannels}} & \mapcell{0.28\textwidth}{Classical stochastic models reproduce many conductance and selectivity observables} & \mapcell{0.30\textwidth}{Perturbations that isolate coherence or resonance predictions without broadly shifting classical energetics}\\\addlinespace

\mapcell{0.15\textwidth}{Microtubules} & \mapcell{0.25\textwidth}{Excited-state transport, collective optical effects, or spin-sensitive signatures may occur in tubulin assemblies \cite{kalra2023electronic,craddock2014feasibility}} & \mapcell{0.28\textwidth}{In vitro signatures, magnetic-isotope perturbation reports, and multiple photophysical models \cite{kakati2024triplet,patwa2024quantum}} & \mapcell{0.28\textwidth}{Preparation artifacts, sample dependence, competing quenching pathways, and structured classical hopping} & \mapcell{0.30\textwidth}{Independent replication with standardized samples and discriminating observables under physiologically relevant or explicitly controlled in vitro conditions}\\\addlinespace

\mapcell{0.15\textwidth}{Genetic code and quantum formalisms} & \mapcell{0.25\textwidth}{Quantum formalisms may describe codon structure or coding patterns, but do not by themselves establish physical quantum processing in living systems \cite{fimmel2018genetic,bashford2008spectroscopy,patel2008towards,balazs2006some}} & \mapcell{0.28\textwidth}{Descriptive mathematical analyses and speculative origin-of-coding frameworks \cite{fimmel2018genetic,bashford2008spectroscopy,balazs2006some}} & \mapcell{0.28\textwidth}{Flexible formalisms; evolutionary, stereochemical, and error-tolerance explanations already account for many code features \cite{gonzalez2019origin,wills2023origins,wills2019reflexivity}} & \mapcell{0.30\textwidth}{Either a measurable quantum signature affecting coding or regulation, or a predictive advantage under strict model comparison and out-of-sample validation}\\\addlinespace

\mapcell{0.15\textwidth}{Posner clusters} & \mapcell{0.25\textwidth}{Protected nuclear spins in Ca$_9$(PO$_4$)$_6$ may influence biochemistry \cite{Fisher2015}} & \mapcell{0.28\textwidth}{Structural plausibility, computational studies, and lifetime constraints \cite{Swift2018,PlayerHore2018}} & \mapcell{0.28\textwidth}{Formation, symmetry, spin relaxation, entanglement generation, and biochemical transduction under physiological conditions} & \mapcell{0.30\textwidth}{In situ detection plus spin-relaxation, coherence, and nonclassical-correlation measurements under physiologically relevant conditions}\\
\bottomrule
\end{tabular}
\end{adjustbox}
\end{table*}
\begin{table*}[t]
\caption{Navigation table for Quantum for Biology. Rows summarize where quantum technology is established, where it is emerging, and which benchmarks would most strongly define progress.}
\label{tab:map_qfb}
\begin{adjustbox}{width=\textwidth}
\begin{tabular}{lllll}
\toprule
\mapcell{0.16\textwidth}{Topic} & \mapcell{0.22\textwidth}{Use case} & \mapcell{0.20\textwidth}{Strongest evidence base} & \mapcell{0.20\textwidth}{Main confounds} & \mapcell{0.16\textwidth}{Decisive benchmarks}\\
\midrule
\mapcell{0.16\textwidth}{Quantum simulation} & \mapcell{0.22\textwidth}{Chemistry and biomolecular simulation beyond some classical scaling regimes \cite{kassal2011simulating,Cao2020}} & \mapcell{0.20\textwidth}{Algorithms and early hardware demonstrations \cite{Bauer2020,navickas2025experimental}} & \mapcell{0.20\textwidth}{Weak baselines and opaque error budgets} & \mapcell{0.16\textwidth}{Blind benchmarks with explicit cost metrics and strong classical comparisons \cite{cordier2022biology,marx2021biology}}\\\addlinespace

\mapcell{0.16\textwidth}{Drug discovery} & \mapcell{0.22\textwidth}{Hybrid workflows for molecular design, binding, and optimization \cite{dong2023prediction}} & \mapcell{0.20\textwidth}{Prototypes and methodological studies \cite{das2024brief,kumar2024recent}} & \mapcell{0.20\textwidth}{Data quality, leakage, and baseline selection} & \mapcell{0.16\textwidth}{Prospective predictions with experimental validation and prespecified metrics \cite{sancho2026npjdrug}}\\\addlinespace

\mapcell{0.16\textwidth}{Quantum machine learning for bioinformatics} & \mapcell{0.22\textwidth}{Omics analysis, similarity search, and optimization \cite{mokhtari2024new,chagneau2024quantum}} & \mapcell{0.20\textwidth}{Reviews, methodological papers, and early demonstrations \cite{maheshwari2022quantum}} & \mapcell{0.20\textwidth}{Leakage, overfitting, and small-dataset effects} & \mapcell{0.16\textwidth}{Curated benchmarks with blind splits and transparent compute reporting}\\\addlinespace

\mapcell{0.16\textwidth}{NV sensing} & \mapcell{0.22\textwidth}{Magnetic and thermal sensing at micro- and nanoscale \cite{Schirhagl2014,Barry2016,degen2017quantum}} & \mapcell{0.20\textwidth}{Replicated sensor physics and biological deployments} & \mapcell{0.20\textwidth}{Calibration, surface effects, and inference gaps} & \mapcell{0.16\textwidth}{Cross-laboratory replication, orthogonal validation, and uncertainty propagation}\\\addlinespace

\mapcell{0.16\textwidth}{Quantum-enhanced NMR/MRI} & \mapcell{0.22\textwidth}{Nano- and microscale NMR and enhanced MR sensitivity \cite{allert2022advances}} & \mapcell{0.20\textwidth}{Rapid methodological development \cite{neuling2023prospects}} & \mapcell{0.20\textwidth}{Resolution and biocompatibility in complex media} & \mapcell{0.16\textwidth}{Benchmark molecular readouts with cross-validation to established assays \cite{yukawa2025qls}}\\\addlinespace

\mapcell{0.16\textwidth}{Quantum imaging} & \mapcell{0.22\textwidth}{Dose-limited imaging and precision enhancement \cite{lemos2014quantum,moreau2019imaging}} & \mapcell{0.20\textwidth}{Biological demonstrations and emerging tissue-scale tests \cite{taylor2013biological,zhang2024quantum}} & \mapcell{0.20\textwidth}{Detector losses and classical noise} & \mapcell{0.16\textwidth}{Dose-matched comparisons against the strongest classical modalities}\\\addlinespace

\mapcell{0.16\textwidth}{Quantum pupillometry and retinal maps} &
\mapcell{0.22\textwidth}{Retinal detection-efficiency maps and pupil-based visual-function or biometric contrast \cite{tinsley2016direct,holmes2017temporal,hecht1942energy,rieke1998single,loulakis2017quantum,margaritakis2020stimulus}} &
\mapcell{0.20\textwidth}{Single-photon visual-detection literature and applied retinal/pupillometric proposals} &
\mapcell{0.20\textwidth}{Physiological variability, attention and adaptation, protocol and device dependence, privacy and security constraints \cite{morales2019irispad,liebling2014privacy,liu2019eyeprivacy}} &
\mapcell{0.16\textwidth}{Large-cohort blinded protocols, robust error-rate estimation, and presentation-attack or privacy models}\\\addlinespace

\mapcell{0.16\textwidth}{Quantum-inspired decision theory} & \mapcell{0.22\textwidth}{Formal models for contextuality, order effects, and interference-like behavior in cognition and decision making \cite{khrennikov2009quantum,khrennikov2015quantum,busemeyer2012quantum,pothos2022quantum}} & \mapcell{0.20\textwidth}{Established mathematical frameworks and representative behavioral fits \cite{pothos2022quantum,wang2014context}} & \mapcell{0.20\textwidth}{Category confusion between formal analogy and physical mechanism; model flexibility} & \mapcell{0.16\textwidth}{Preregistered out-of-sample prediction and strict model comparison that penalizes flexibility}\\
\bottomrule
\end{tabular}
\end{adjustbox}
\end{table*}

\begin{table*}[t]
\caption{Navigation table for Biology for Quantum. Rows summarize biological contributions that can be evaluated by device- or algorithm-level performance.}
\label{tab:map_bfq}
\begin{adjustbox}{width=\textwidth}
\begin{tabular}{lllll}
\toprule
\mapcell{0.18\textwidth}{Topic} & \mapcell{0.28\textwidth}{Central idea} & \mapcell{0.20\textwidth}{Strongest evidence base} & \mapcell{0.20\textwidth}{Main uncertainty} & \mapcell{0.20\textwidth}{   Benchmarks   }\\
\midrule
\mapcell{0.18\textwidth}{DNA origami scaffolds} & \mapcell{0.28\textwidth}{Programmable self-assembly positions emitters and nanostructures with nanometer-scale precision \cite{kuzyk2018dnaorigami,hong2017dnaorigami,zhan2023dnaorigami}} & \mapcell{0.20\textwidth}{Robust nanoscale assembly demonstrations} & \mapcell{0.20\textwidth}{Stability, yield, and integration in device settings} & \mapcell{0.20\textwidth}{Reproducible coupling distributions and integration with photonics}\\\addlinespace

\mapcell{0.18\textwidth}{Virus-templated materials} & \mapcell{0.28\textwidth}{Biological templates guide synthesis and assembly of nanowires and related nanostructures \cite{mao2004virusnanowires}} & \mapcell{0.20\textwidth}{Controlled synthesis and assembly in vitro} & \mapcell{0.20\textwidth}{Whether loss and disorder can meet quantum-device requirements} & \mapcell{0.20\textwidth}{Performance distributions benchmarked against top-down fabrication}\\\addlinespace

\mapcell{0.18\textwidth}{DNA-templated wiring} & \mapcell{0.28\textwidth}{Biological scaffolds define conductive nanoscale pathways \cite{braun1998dnaSilverWire}} & \mapcell{0.20\textwidth}{Electrically connected nanoscale structures} & \mapcell{0.20\textwidth}{Interface quality and purity under quantum-device conditions} & \mapcell{0.20\textwidth}{Loss, noise, and reproducibility under device conditions}\\\addlinespace

\mapcell{0.18\textwidth}{Biohybrid polaritons} & \mapcell{0.28\textwidth}{Biological chromophore networks can strongly couple to optical cavities \cite{coles2014chlorosomes}} & \mapcell{0.20\textwidth}{Demonstrated strong coupling in biohybrid systems} & \mapcell{0.20\textwidth}{Stability, controllability, and scalability} & \mapcell{0.20\textwidth}{Reproducible coupling, tunability, and circuit integration}\\\addlinespace

\mapcell{0.18\textwidth}{Genetically encoded qubits} & \mapcell{0.28\textwidth}{Biomolecules can host controllable quantum states \cite{feder2025fluorescentqubit}} & \mapcell{0.20\textwidth}{Demonstrated qubit platform in a protein} & \mapcell{0.20\textwidth}{Fidelity and coherence in complex environments} & \mapcell{0.20\textwidth}{Standard qubit metrics and reproducibility across expression contexts}\\\addlinespace

\mapcell{0.18\textwidth}{Bioinspired transport} & \mapcell{0.28\textwidth}{Noise-assisted transport and interference-inspired design principles motivated by photosynthesis \cite{plenio2008dephasing,Creatore2013}} & \mapcell{0.20\textwidth}{Theory and early prototypes \cite{Romero2014}} & \mapcell{0.20\textwidth}{Robustness to disorder and scalability} & \mapcell{0.20\textwidth}{Device gains that persist across disorder ensembles}\\\addlinespace

\mapcell{0.18\textwidth}{Quantum biomimetics and artificial-life-inspired algorithms} & 
\mapcell{0.28\textwidth}{Life-like behaviors such as self-replication, mutation, interaction, inheritance, adaptation, and death can be used as design motifs for quantum-information protocols and controllable artificial-life models \cite{AlvarezRodriguez2016ArtificialLifeQT,AlvarezRodriguez2018QuantumArtificialLife,Valente2021SelfReplication}} & 
\mapcell{0.20\textwidth}{Quantum artificial-life models, biomimetic cloning protocols, and proof-of-principle implementation on an IBM quantum computer \cite{AlvarezRodriguez2014BiomimeticCloning,AlvarezRodriguez2018QuantumArtificialLife,Lamata2020QMLBiomimetics}} & 
\mapcell{0.20\textwidth}{Currently exploratory; unclear advantage over classical artificial-life models, evolutionary algorithms, or standard quantum algorithms} & 
\mapcell{0.20\textwidth}{Matched benchmarks testing scalability, resource use, robustness, and task-level advantage against classical artificial-life and standard quantum baselines}\\\addlinespace

\mapcell{0.18\textwidth}{Brain-inspired quantum and quantum-inspired neural computation} & \mapcell{0.28\textwidth}{Brain-inspired architectures motivate both quantum neuromorphic designs and quantum-like models for complex data, cognition, and neuroimaging \cite{Kudithipudi2025,markovic2020quantum,brand2024quantum}} & \mapcell{0.20\textwidth}{Classical neuromorphic hardware, proof-of-principle quantum neuromorphic proposals, and quantum-like modeling frameworks \cite{Kudithipudi2025,brand2024quantum,Scholes2024QLNetworks,KhrennikovYamada2025MentalEntanglement}} & \mapcell{0.20\textwidth}{Whether such approaches outperform strong classical neuromorphic, deep-learning, or standard quantum baselines on well-defined tasks} & \mapcell{0.20\textwidth}{Matched benchmarks for scalability, robustness, resource demands, and generalization against strong classical and standard quantum baselines}\\
\bottomrule
\end{tabular}
\end{adjustbox}
\end{table*}
\clearpage
\bibliographystyle{apsrev4-2}
\bibliography{biblio-qbioreview_ordered}

\end{document}